\documentclass[apj]{emulateapj}
\usepackage{graphicx}

\def\simlt{\mathrel{\hbox{\rlap{\hbox{\lower4pt\hbox{$\sim$}}}\hbox{$<$}}}}
\def\simgt{\mathrel{\hbox{\rlap{\hbox{\lower4pt\hbox{$\sim$}}}\hbox{$>$}}}}
\newcommand{\etal}{et al.}

\def\ro{{\it ROSAT\/}}
\def\chandra{{\it Chandra}}
\def\xmm{{\it XMM--Newton}}
\def\pks{{\rm PKS~1209$-$51/52}}

\def\epsr{{\rm PSR~J1210$-$5226}}

\def\psr{\rm{PSR J1852$+$0040}}
\def\snr{Kes~79}
\def\psnr{Puppis~A}
\def\pup{RX~J0822$-$4300}
\def\puppsr{\rm{PSR~J0821$-$4300}}
\def\cal{1RXS J141256.0$+$792204}

\slugcomment{}

\shorttitle{Spin-down of \puppsr\ in \psnr}
\shortauthors{Gotthelf et al.}

\begin{document}

\title{The Spin-down of PSR J0821--4300 and PSR J1210--5226: \\
Confirmation of Central Compact Objects as Anti-Magnetars}

\author{E. V. Gotthelf, J. P. Halpern, and J. Alford}

\affil{Columbia Astrophysics Laboratory, Columbia University, 550 West 120th Street, New York, NY 10027}

\begin{abstract}
Using \xmm\ and \chandra, we measure period derivatives for
the second and third known pulsars in the class of
Central Compact Objects (CCOs) in supernova remnants,
proving that these young neutron stars have exceptionally
weak dipole magnetic field components. For the 112~ms \puppsr\ in \psnr,
$\dot P=(9.28\pm0.36)\times10^{-18}$.
Its proper motion, $\mu=61\pm9$~mas~yr$^{-1}$,
was also measured using \chandra.
This contributes a kinematic term to the period derivative
via the Shklovskii effect, which is subtracted from $\dot P$ to derive dipole
$B_s=2.9\times10^{10}$~G, a value similar to that of first
measured CCO \psr\ in \snr, which has $B_s=3.1\times10^{10}$~G.
Antipodal surface hot spots with different temperatures and
areas are deduced from the X-ray spectrum and
pulse profiles. Paradoxically, such nonuniform surface temperature
appears to require strong crustal magnetic fields,
probably toroidal or quadrupolar components
much stronger than the external dipole.
A spectral feature, consisting of
either an emission line at $\approx0.75$~keV or absorption
at $\approx0.46$~keV, is modulated in strength with
the rotation. It may be due to a cyclotron process
in a magnetic field on the surface that is slightly
stronger than the dipole deduced from the spin-down.
We also timed anew the 424~ms \epsr, resolving previous ambiguities
about its spin-down rate.  Its $\dot P = (2.22\pm0.02)\times10^{-17}$,
corresponding to $B_s=9.8\times10^{10}$~G.  This is
compatible with a cyclotron resonance interpretation of its prominent
absorption line at 0.7~keV and harmonics.  These results
deepen the mystery of the origin and evolution of CCOs:
why are their numerous descendants not evident?
\end{abstract}


\section {Introduction}

The class of faint X-ray sources in supernova remnants (SNRs)
known as central compact objects (CCOs) are characterized by
steady flux, predominantly surface thermal X-ray emission,
lack of a surrounding pulsar wind nebula, and absence of
detection at any other wavelength.  Table~\ref{ccos} lists basic data
on the well-studied CCOs, as well as proposed candidates whose
qualifications are not yet well established.  Of the eight most
secure CCOs, three are known to be neutron stars (NSs) with spin
periods of 0.105, 0.424, and 0.112~s.  Spin-down has
been detected for two of these, the 0.105~s pulsar \psr\
in Kes~79 and the 0.424~s pulsar \epsr\ in the SNR \pks.
For \psr, the implied surface dipole field is only
$B_s = 3.1 \times 10^{10}$~G \citep{hal10a}, smaller than that
of any other young known NS.  In the case of \epsr, archival
data allow two alternative timing solutions, with
$B_s = 9.9 \times 10^{10}$ or $2.4 \times 10^{11}$~G
\citep{hal11a}. 

It is natural to assume that CCOs that have not yet been
seen to pulse are isolated, weakly magnetized NSs of the
same class as the CCO pulsars. Where pulsar searches have been
unsuccessful, it is possible that an even weaker magnetic field,
a more uniform surface temperature, or an unfavorable viewing
geometry, prevents detection of rotational modulation.
The absence of pulsations from the youngest known NS,
the $\approx 330$ year old CCO in Cassiopeia~A, has been used,
in combination with fitting of its X-ray spectrum, to argue that
it is covered with a uniform temperature, non-magnetized atmosphere
of carbon, the product of nuclear burning of H and He \citep{ho09}.
Rapid cooling of the NS in Cas~A, directly detected by \chandra\
\citep{hei10}, has been interpreted as evidence 
for neutron superfluidity in the core \citep{pag11,sht11}.

\begin{deluxetable*}{llcccrccl}[hb]
\tabletypesize{\scriptsize}
\tablewidth{0pt}
\tablecaption{Central Compact Objects in Supernova Remnants}
\tablehead{
\colhead{CCO} & \colhead{SNR} & \colhead{Age} & \colhead{$d$} & \colhead{$P$} &
\colhead{$f_p$\tablenotemark{a}} & \colhead{$B_s$} & \colhead{$L_{x,\rm bol}$} & \colhead{References} \\
\colhead{} & \colhead{} & \colhead{(kyr)} & \colhead{(kpc)} & \colhead{(s)} & \colhead{(\%)}
 &
\colhead{($10^{10}$~G)} & \colhead{(erg~s$^{-1}$)}
}
\startdata
RX~J0822.0$-$4300        & Puppis~A         & 4.5        & 2.2   & 0.112   & 11     & 2.9     & $5.6 \times 10^{33}$    & 1,2,3,4,5,6 \\
CXOU~J085201.4$-$461753  & G266.1$-$1.2     & 1          & 1     & \dots\  & $<7$   & \dots\  & $2.5 \times 10^{32}$    & 7,8,9,10,11 \\
1E 1207.4$-$5209         & PKS~1209$-$51/52 & 7          & 2.2   & 0.424   & 9      & 9.8     & $2.5 \times 10^{33}$    & 6,12,13,14,15,16,17 \\
CXOU~J160103.1$-$513353  & G330.2$+$1.0     & $\simgt 3$ & 5     & \dots\  & $<40$  & \dots\  & $1.5 \times 10^{33}$    & 18,19 \\
1WGA~J1713.4$-$3949      & G347.3$-$0.5     & 1.6        & 1.3   & \dots\  & $<7$   & \dots\  & $\sim 1 \times 10^{33}$ & 11,20,21  \\
XMMU~J172054.5$-$372652  & G350.1$-$0.3     & 0.9        & 4.5   & \dots\  & \dots\ & \dots\  & $3.9 \times 10^{33}$    & 22,23 \\
CXOU~J185238.6$+$004020  & Kes~79           & 7          & 7     & 0.105   & 64     & 3.1     & $5.3 \times 10^{33}$    & 24,25,26,27 \\
CXOU~J232327.9$+$584842  & Cas~A            & 0.33       & 3.4   & \dots\  & $<12$  & \dots\  & $4.7 \times 10^{33}$    & 27,28,29,30,31,32,33 \\
\hline
2XMMi~J115836.1$-$623516 & G296.8$-$0.3     & 10         & 9.6   & \dots\  & \dots\ & \dots\  & $1.1\times 10^{33}$     & 34 \\
XMMU~J173203.3$-$344518  & G353.6$-$0.7     & $\sim 27$  & 3.2   & \dots\  & $<9$   & \dots\  & $1.3 \times 10^{34}$    & 35,36,37,38 \\
CXOU~J181852.0$-$150213  & G15.9$+$0.2      & $1-3$      & (8.5) & \dots\  & \dots\ & \dots\  & $\sim 1 \times 10^{33}$ & 39
\enddata
\tablecomments{Above the line are eight well-established CCOs.
Below the line are three candidates.}
\tablenotetext{a}{Upper limits on pulsed fraction are for a search down to $P=12$~ms or smaller.}
\tablerefs{
(1) \citealt{hui06a};
(2) \citealt{got09};
(3) \citealt{got10};
(4) \citealt{del12};
(5) \citealt{bec12};
(6) this paper;
(7) \citealt{sla01};
(8) \citealt{kar02};
(9) \citealt{bam05};
(10) \citealt{iyu05};
(11) \citealt{del08};
(12) \citealt{zav00};
(13) \citealt{mer02a};
(14) \citealt{big03};
(15) \citealt{del04};
(16) \citealt{got07};
(17) \citealt{hal11a};
(18) \citealt{par06};
(19) \citealt{par09};
(20) \citealt{laz03};
(21) \citealt{cas04};
(22) \citealt{gae08};
(23) \citealt{lov11};
(24) \citealt{sew03};
(25) \citealt{got05};
(26) \citealt{hal07};
(27) \citealt{hal10a};
(28) \citealt{pav00};
(29) \citealt{cha01};
(30) \citealt{mer02b};
(31) \citealt{pav09};
(32) \citealt{ho09};
(33) \citealt{hei10};
(34) \citealt{san12};
(35) \citealt{tia08};
(36) \citealt{abr11};
(37) \citealt{hal10b};
(38) \citealt{hal10c};
(39) \citealt{rey06}.
}
\label{ccos}
\end{deluxetable*}

The ``anti-magnetar'' explanation of CCOs, which is motivated by their
weak magnetic fields, absence of variability, and location on the
$P-\dot P$ diagram, remains incomplete in detail.
Specifically, it does not yet account for
the hot spots that are seen on the surfaces of CCO pulsars.
Since the spin-down power of a CCO pulsar is
less than its X-ray luminosity,
the latter must be thermal emission from
residual cooling, which can only be nonuniform if
there is anisotropic heat conduction.  In the absence of
strong magnetic fields or magnetospheric activity, it is difficult
to reproduce the light curve and
pulsed fraction of 64\% from \psr\ in \snr\ \citep{hal10a,sha12},
or the two antipodal hot spots of different temperatures
and areas on \pup\ in Puppis~A \citep{got09,got10}.
The latter 0.112~s pulsar, hereafter \puppsr, is a
subject of this paper.  Its spectrum is especially
puzzling in also displaying a phase-dependent
emission feature at 0.7--0.8~keV \citep{got09},
which is reported to be variable in the long term 
\citep{del12}.

Here we report the first spin-down measurement for \puppsr,
based on a dedicated program of phase-coherent timing
jointly scheduled between \xmm\ and \chandra. 
It was also necessary to incorporate \chandra\ HRC
observations of the position and proper motion of \puppsr\
in order to determine its small period derivative accurately.
The astrometric analysis is described in Section 2.
(The latter work was also carried out, with consistent results,
by \citealt{bec12}.)
The results of the timing are presented
in Section 3.  The X-ray flux and spectra are discussed in Section 4,
with particular attention paid to the spectral line and the
question of its possible variability.
We also obtained new timing observations of \epsr\
that resolve the prior ambiguity about its spin-down
rate in favor of the smaller value;
this definitive result is presented in Section 5.
The nature of CCOs as anti-magnetars, and their possible
evolutionary status, are discussed in Section 6.
Conclusions and proposals for future work follow in
Section 7.

\section{X-ray Position and Proper Motion}

Evidence that \puppsr\ has high proper motion from \chandra\
HRC images over a 5 year baseline was reported by \citet{hui06b}
and \citet{win07}, but with slightly disparate measurements of
$\mu = 107\pm 34$ mas~yr$^{-1}$ and $\mu = 165\pm 25$ mas~yr$^{-1}$,
respectively, from the same data.  Here, we are concerned
with timing this high-velocity pulsar over an extended period of time
with millisecond accuracy.  When trying to measure a small $\dot P$
using X-ray data, position and proper motion can contribute significant
errors via three effects.   The first is an instrumental property of the
\chandra\ CCDs when used in continuous-clocking (CC) mode (see Section 3);
the position of the pulsar must be known a priori to
$< 0.\!^{\prime\prime}5$ in order to determine the
time of arrival of each source photon.  The second consideration
is the accuracy of the barycentric correction.  The third effect
is the magnitude of the proper motion, which contributes a purely
kinematic period derivative via the ``train whistle'' effect \citep{shk70}.
The original measurements of proper motion are not accurate enough to
measure this effect, which is crucial in the case of \puppsr.

Accordingly, we have reanalyzed the position and proper motion of \puppsr\  
using the \chandra\ HRC-I data listed in Table~\ref{hrclog}, which now
includes a more recent pair of observations in 2010 August that extends
the baseline to 10.6 yr, enabling higher precision on both the
contemporary position for timing, and the proper motion.  We will
describe here any differences between our method and previous work.
For example, we did not use an HRC-S observation (ObsID 1851) because
of known systematic differences between HRC-S and HRC-I.
Ultimately, however, our results are consistent with the 
recent analysis of the same data (including ObsID 1851) by \citet{bec12}.

\begin{deluxetable*}{ccccccc}
\tabletypesize{\small}
\tablewidth{0pt}
\tablecaption{Log of \chandra\ HRC-I Observations of \puppsr}
\tablehead{
\colhead{ObsID} & \colhead{Date} & \colhead{Start Epoch} & \colhead{Exposure} & \colhead{Roll angle} & \colhead{Star A}& \colhead{\puppsr} \\
\colhead{}  & \colhead{(UT)} & \colhead{(MJD)} & \colhead{(ks)} & \colhead{($^{\circ}$)} & \colhead{(Counts)\tablenotemark{a}} & \colhead{(Counts)\tablenotemark{a}}
}    
\startdata
749     & 1999 Dec 21 &   51533.95  & 18.0 & 338.7  & 47  & 3257 \\
4612    & 2005 Apr 25 &   53485.31  & 40.2 & 261.9  & 123 & 7260 \\
11819   & 2010 Aug 10 &   55418.72  & 33.7 & 163.4  & 101 & 5455 \\
12201   & 2010 Aug 11 &   55419.13  & 38.7 & 162.9  & 117 & 6296
\enddata
\tablenotetext{a}{Total counts collected in a $1.\!^{\prime\prime}5$
radius aperture centered on the source.}
\label{hrclog}
\end{deluxetable*}

The data from all epochs were reprocessed and analyzed using the latest
calibration files and software (CIAO 4.4/CALDB 4.4.8). This processing
accounts for the HRC {\tt AMP\_SF} electronic ringing distortions discussed
by \cite{hui06b}.
The HRC detector is well suited to astrometry, with its processed
pixel size of $0.\!^{\prime\prime}1318$ that oversamples the on-axis point spread
function (PSF) by a factor of 5.  For all observations, the pulsar was
placed close to the optical axis where the PSF is essentially symmetric.
In the following analysis we assume, as there is no evidence to the contrary,
that the HRC focal plane is linear and the aspect reconstruction introduces
no errors in roll angle.  The two pointings on consecutive days
in 2010 August are sufficiently different in their aspect reconstruction
that we analyze them individually.

The nominal uncertainty in aspect reconstruction for a typical \chandra\
observation is $0.\!^{\prime\prime}6$.  It is often possible to remove most
of this systematic error by using nearby X-ray point sources with
precisely measured optical coordinates to correct the absolute astrometry.
\cite{hui06b} used
their X-ray detected ``star~A'' as their sole fiducial point,
and fitted a model PSF interpolated from the CIAO library appropriate
for its position and estimated photon energy to determine its position.
\cite{win07} used this star and two additional stars, and followed
the updated method outlined in the CIAO thread for generating a Monte
Carlo PSF using the {\tt CHaRT/MARX} software for input into the
CIAO/{Sherpa} spectral/image fitting software package.

\begin{figure*}
\centerline{
\hfill
\includegraphics[angle=270,width=0.28\linewidth,clip=]{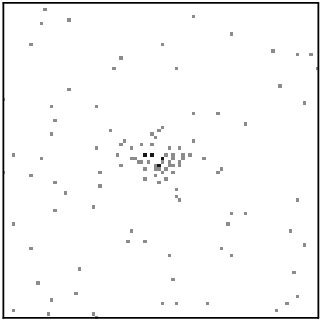}
\hfill
\includegraphics[angle=270,width=0.28\linewidth,clip=]{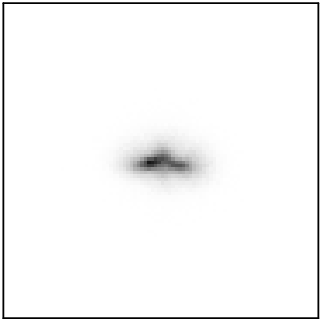}
\hfill
}
\centerline{
\hfill
\includegraphics[angle=270,width=0.28\linewidth,clip=]{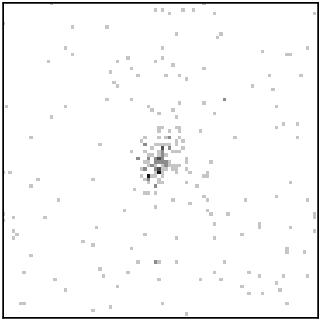}
\hfill
\includegraphics[angle=270,width=0.28\linewidth,clip=]{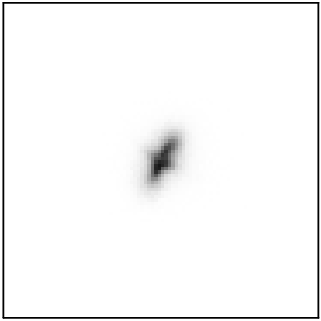}
\hfill
}
\centerline{
\hfill
\includegraphics[angle=270,width=0.28\linewidth,clip=]{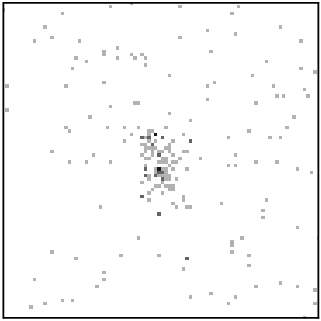}
\hfill
\includegraphics[angle=270,width=0.28\linewidth,clip=]{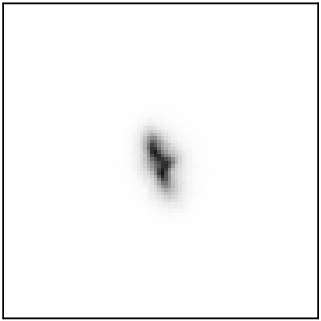}
\hfill
}
\centerline{
\hfill
\includegraphics[angle=270,width=0.28\linewidth,clip=]{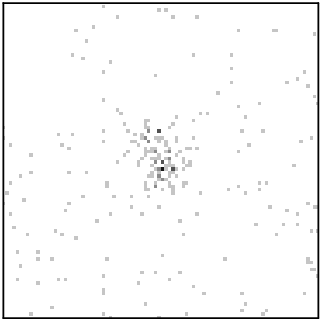}
\hfill
\includegraphics[angle=270,width=0.28\linewidth,clip=]{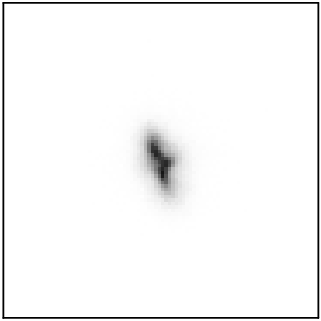}
\hfill
}
\caption{\chandra\ HRC-I images around reference star~A (3UC094-058669) 
used to define the coordinate system for position and proper motion
of \puppsr.  Each panel shows the observed counts from Star~A (left)
and the simulated PSF (right) for its location on the focal plane in
native HRC-I pixels. The plots cover $12^{\prime\prime}\times 12^{\prime\prime}$
in celestial coordinates. The total counts in a $1.\!^{\prime\prime}5$
radius aperture are given in Table~\ref{coordtable}.}
\label{starApsf}
\end{figure*}

In our analysis, we also follow the CIAO thread to characterize
the HRC-I PSF, but we adopt a simpler approach to measuring source
locations, one that is not dependent on model fitting in the image
domain.  Our method is guided by the following observations.
First, the on-axis, symmetric image of the pulsar contains enough
counts that a simple centroid calculation is a sufficiently accurate
measurement of its position on the detector.  Second, star~A of 
\citet{hui06b}, which lies $2\farcm7$ from \puppsr, is the only
useful fiducial source for registering the X-ray image.
The position and proper motion of star~A are taken from the
UCAC3 \citep{zac10}, where it is listed as 3UC094-058669 with
coordinates (J2000.0) R.A.=$08^{\rm h}21^{\rm m}46.\!^{\rm s}2924(16)$,
decl.=$-43^{\circ}02^{\prime}03.\!^{\prime\prime}640(49)$, and
proper motion $\mu_\alpha\,{\rm cos}\,\delta = -14.3(2.0)$,
$\mu_\delta = -3.6(5.5)$ mas~yr$^{-1}$.
X-ray position measurements of the two weaker, off-axis stars used
by \citet{win07} only add to the uncertainty (as quantified below)
in the absolute astrometry.  Third, the few X-ray photons from
star~A, and its broad off-axis PSF, do not warrant a sophisticated
image fitting technique.  Instead, we use a ``corrected centroid''
method, as described below. 

To determine the source location of star~A in the X-ray images we
start with the {\tt CHaRT/MARX} simulation of the PSF,
as described in the CIAO user
webpages\footnote{http://cxc.harvard.edu/chart},
for its respective locations on the focal plane.
Figure~\ref{starApsf}
shows the distribution of the counts from each HRC image and
corresponding Monte Carlo PSF. It is apparent that
star~A is poorly sampled in the data, with total source counts
in the range $47 \leq N \leq 123$ (Table~\ref{hrclog}). The maximum number of
counts per pixel is typically only $2-4$, making forward
fitting poorly constrained statistically, while sharp features in the model
can cause systematic offsets. Furthermore, the source is immersed in a
substantial diffuse background from the \psnr\ SNR which,
although it only contributes a few photons over the source region,
adds uncertainty to the position measurement because of the small
count statistics, especially when fitting over a larger area.

\begin{figure}
\centerline{
\hfill
\includegraphics[angle=270,width=0.9\linewidth,clip=]{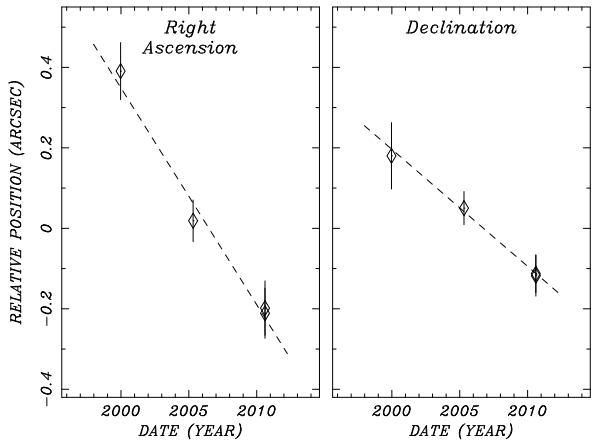}
\hfill
}
\caption{Four position measurements of \puppsr\ spanning 10.6 yr, after
correction using the optical/X-ray star~A (3UC094-058669) as a
reference.  The two observations in 2010 nearly coincide in time.
The positions (diamonds) and their errors are fitted with a linear model
(dashed lines). The fitted parameters are listed in Table~\ref{ephemeris}.}
\label{pmfit}
\end{figure}

To sidestep these effects,
the coordinates of star~A are first found from its
centroid. Photons were extracted from circular aperture of radius
$1.\!^{\prime\prime}5$, chosen to minimize the background counts. This
aperture encloses essentially all of the signal in that fraction of
the PSF with a finite probability of producing a single count during
the observation.  This measurement was made using the CIAO tool {\it
dmstat} and is iterated to produce the final coordinates. However,
while this results in a well-defined and statistically meaningful
measurement, it does not account for the shape of the complex off-axis
PSF, whose orientation depends on the spacecraft roll angle (which
differs for each observation; see Figure~\ref{starApsf}), or for the
monotonic off-axis deviations introduced by the flat focal plane. To
quantify these systematic effects we also measure
the centroid of the simulated PSF of star~A in sky coordinates in each
image and compare it to the coordinates that were input to
{\tt CHaRT/MARX} for that PSF. The difference constitutes the small,
but critical correction to the centroid of star~A.

We estimated the uncertainty in the derived coordinates of star~A
using a Monte Carlo method. We generated 500 realizations of star~A by
sampling the PSF using a random number generator to match the observed
counts, and accumulated the centroid measurements to build up a
distribution in right ascension and declination.  To account for the
observed background, we included a random distribution of photons
within the source aperture.  The resulting (Gaussian) width of the
distribution of centroids is typically $\sigma \approx
0.\!^{\prime\prime}06$ in each of the two coordinates (see
Table~\ref{coordtable}). These reproduce the expected ``standard
error'' for a centroid, $\approx{\sigma}/\sqrt{N}$.  We also simulated
the uncertainties for the two other fiducial sources used in
\cite{win07} and find that their inclusion in the position
determination would only increase the uncertainty on the final
coordinates.

The pulsar itself is a strong source (see Table~\ref{hrclog}) whose
coordinates are precisely measured using a centroid calculation with
uncertainty an order of magnitude smaller than those measured for
star~A.  Because of the symmetry of its on-axis PSF, no systematic
correction is required for the pulsar.  Its coordinates are then
adjusted by the difference between the optical and X-ray coordinates of
star~A at each epoch to produce its final astrometric position
in Table~\ref{coordtable}.  Fitting the position of the
pulsar as a function of time then yields the proper motion.
Figure~\ref{pmfit} shows the $\chi^2$ fit with constant velocity
in each coordinate,  and Table~\ref{ephemeris} lists the
formal solution and quantities derived from it.
The derived total proper motion of $61\pm 9$ mas~yr$^{-1}$ is in agreement
with the value determined by \citet{bec12}, $71\pm 12$ mas~yr$^{-1}$,
and is smaller than
previously published numbers for reasons discussed in that paper.

\begin{deluxetable*}{ccccccc}
\tabletypesize{\scriptsize}
\tablewidth{0pt}
\tablecaption{Position Measurements for \puppsr}
\tablehead{
\colhead{Epoch} & \multicolumn{2}{c}{Star~A (Optical)}
& \multicolumn{2}{c}{Star~A (X-ray)} & \multicolumn{2}{c}{\puppsr\ (corrected)} \\
\colhead{(year)} &
\colhead{R.A. (h m s)} &
\colhead{Decl. ($^{\circ}\ ^{\prime}\ ^{\prime\prime}$)} &
\colhead{R.A. (h m s)} &
\colhead{Decl. ($^{\circ}\ ^{\prime}\ ^{\prime\prime}$)} &
\colhead{R.A. (h m s)} &
\colhead{Decl. ($^{\circ}\ ^{\prime}\ ^{\prime\prime}$)}
}
\startdata
1999.975    &   08 21 46.2928(16)  & --43 02 03.637(49)  & 08 21 46.2882(65)  & -43 02 03.397(83)  & 08 21 57.4024(67)  & -43 00 16.894(96)\\  
2005.316    &   08 21 46.2858(21)  & --43 02 03.657(57)  & 08 21 46.3011(47)  & -43 02 03.756(41)  & 08 21 57.3685(52)  & -43 00 17.023(71)\\ 
2010.610   &   08 21 46.2789(31)  & --43 02 03.676(76)  & 08 21 46.2621(62)  & -43 02 04.113(47)  & 08 21 57.3488(69)  & -43 00 17.185(89)\\ 
2010.611   &   08 21 46.2789(31)  & --43 02 03.676(76)  & 08 21 46.2575(57)  & -43 02 04.068(51)  & 08 21 57.3476(65)  & -43 00 17.190(92)
\enddata									 
\tablecomments{All coordinates are equinox J2000.  Optical coordinates for star~A
(3UC094-058669) are corrected for the epoch of proper motion.
The X-ray position of star~A is determined using the method described
in the text.  The pulsar coordinates are corrected by the difference between
the optical and X-ray coordinates of star~A. 
Uncertainties on the last digits are in parentheses.}
\label{coordtable}
\end{deluxetable*}

The tangential velocity of \puppsr\ then depends on a distance
determination for \psnr,
which has ranged from 1 to 2.5~kpc according to various methods,
and is a matter of unresolved debate in the most
recent studies quoted here.  One \citep{rey95,rgj03}
is based on 21~cm \ion{H}{1} velocities, and a perceived morphological
association of features in \ion{H}{1} surrounding the pulsar
and the SNR at $v_{\rm lsr}=+16$~km~s$^{-1}$, the latter corresponding
to $d=2.2\pm 0.3$~kpc.  The second method uses spectra of ground-state
hydroxyl lines \citep{woe00}, which show absorption at
$v_{\rm lsr}<+7.6$~km~s$^{-1}$, and emission above this velocity,
from which $d=1.3^{+0.6}_{-0.8}$~kpc is derived.  Both sets of
authors employ assumptions that are not mutually accepted,
and are beyond the scope of this paper to evaluate.
We will adopt a fiducial distance of $2.2\pm 0.3$~kpc for purposes of further
calculations, while noting the implications of a possible
smaller distance where relevant.

\begin{deluxetable}{lc}
\tablecaption{Ephemeris of \puppsr}
\tablehead{
\colhead{Parameter} & \colhead{Value}
}
\startdata
\multispan{2}{\hfill Position and Proper Motion \hfill} \\
\multispan{2}{\vspace{4pt}} \\
\hline
\multispan{2}{\vspace{4pt}} \\
Epoch of position and $\mu$ (MJD)     & 53964.0 \\
R.A. (J2000)                                  & $08^{\rm h}21^{\rm m}57.\!^{\rm s}3653(31)$ \\
Decl. (J2000)                                 & $-43^{\circ}00^{\prime}17.\!^{\prime\prime}074(43)$ \\
R.A. proper motion, $\mu_{\alpha}\,{\rm cos}\,\delta$  & $-54.1\pm 8.3$ mas yr$^{-1}$ \\
Decl. proper motion, $\mu_{\delta}$             & $-28.1\pm 10.5$ mas yr$^{-1}$ \\
Total proper motion, $\mu$                    & $61.0\pm 8.8$ mas yr$^{-1}$ \\
Position angle of proper motion               & $242.\!^{\circ}5\pm 9.\!^{\circ}5$ \\
Tangential velocity\tablenotemark{a}, $v_{\perp,c}$  & $629\pm 126$ km s$^{-1}$ \\
\cutinhead{Timing Solution}
Epoch of ephemeris (MJD TDB)\tablenotemark{b}     & 55580.0000006 \\
Span of ephemeris (MJD)                       & 55,182--56,027 \\
Frequency, $f$                                & 8.86529105448(32) Hz \\
Frequency derivative, $\dot f$                & $(-7.29 \pm 0.28) \times 10^{-16}$ Hz s$^{-1}$ \\
Period, $P$                                   & 0.1127994550720(41) s \\
Period derivative, $\dot P$                   & $(9.28 \pm 0.36) \times 10^{-18}$ \\
Kinematic period derivative\tablenotemark{a}, $\dot P_k$       & $(2.24\pm 0.72) \times 10^{-18}$ \\
Intrinsic period derivative\tablenotemark{a}, $\dot P_{\rm int}$ & $(7.04\pm 0.80) \times 10^{-18}$ \\
Surface dipole magnetic field, $B_s$          & $2.9 \times 10^{10}$ G\\
Spin-down luminosity, $\dot E$                & $1.9 \times 10^{32}$ erg s$^{-1}$ \\
Characteristic age, $\tau_c$                  & 254 Myr
\enddata
\tablenotetext{a}{Assuming $d=2.2\pm 0.3$ kpc \citep{rey95}.}
\tablenotetext{b}{Epoch of fitted minima of the $1.5-4.5$~keV pulse profile; phase zero in Figure~\ref{lightcurve}.}
\label{ephemeris}
\end{deluxetable}

The tangential velocity of \puppsr\ at $d=2.2\pm 0.3$~kpc
is $v_{\perp} = 636 \pm 126$ km~s$^{-1}$.
This velocity is at the high end of
the distribution of two-dimensional velocities of pulsars measured
by \citet{hob05}, who find mean values of
$\bar v_{\perp}=246\pm 22$~km~s$^{-1}$
for 121 ordinary (non-recycled) pulsars, and 
$\bar v_{\perp}=307\pm 47$~km~s$^{-1}$ for 46 pulsars whose
characteristic ages are $<3$~Myr.  The individual pulsar velocities
are corrected for the solar motion with respect to the local standard
of rest and a flat Galactic rotation curve in order to express them
in the frame of the rotating Galactic disk.  In the case of \puppsr\
this correction is dominated by the solar motion for the range of
plausible distances, and is only $-7$ km~s$^{-1}$, resulting in a
corrected tangential velocity of
$v_{\perp,c} = 629$ km~s$^{-1}$.
If the distance is as small as 1~kpc, this reduces to
an unexceptional 290~km~s$^{-1}$.

The \citet{shk70} effect,
a purely kinematic contribution to the observed period
derivative, will be significant.  For a source
moving at constant velocity the kinematic contribution is
$$\dot P_k\,={\mu^2\,P\,d \over c}\ =\ {v^2_{\perp}\,P \over d\,c}\ .\eqno(1)$$
From the proper motion measurement of \puppsr, 
$\dot P_k = (2.24\pm 0.72)\times 10^{-18}$
is calculated, where we have propagated the uncertainties on both 
$\mu$ and $d$.

\section{X-ray Timing}

Previous observations of \puppsr\ were only
able to set upper limits on its period derivative \citep{got09,got10,del12}.
Evidently $\dot P$ is so small that it can only be measured
by phase-coherent timing.  Accordingly, we designed a
sequence of observations coordinated between \xmm\ and \chandra\
that would start and maintain phase connection over a 2~year span,
2010 May -- 2012 April.  The scheduling strategy is the same
as was used for \psr\ in \snr\ \citep{hal10a}.
By design, the resulting ephemeris was also connected backward
to archival observations that were obtained in 2009 December and 2010 April
\citep{del12}, which extended the time span to 2.3~years.
All of the timing observations used in this analysis are listed
in Table~\ref{timinglog}. The discovery observations from 2001
\citep{got09} are not included, as they are too far removed 
in time to be reliably connected.

\begin{deluxetable*}{llrlcccc}
\tabletypesize{\scriptsize}
\tablewidth{0pt}
\tablecaption{Log of X-ray Timing Observations of \puppsr}
\tablehead{
\colhead{Mission} & \colhead{Instr/Mode} & \colhead{ObsID} & 
\colhead{Date} &
\colhead{Elapsed time/} & \colhead{Start Epoch} & 
\colhead{Period\tablenotemark{a}} & \colhead{$Z^2_1$} \\
\colhead{} & \colhead{} & \colhead{} & \colhead{(UT)} & 
\colhead{Livetime (ks)} & \colhead{(MJD)} &
\colhead{(s)} & \colhead{}
}
\startdata
{\it XMM} & EPIC-pn/SW &0606280101 & 2009 Dec 17,18& 85.1/54.9  &  
55182.820 & 0.112799488(12) &  173.0 \\
{\it XMM} & EPIC-pn/SW &0606280201 & 2010 Apr 05 & 42.2/29.4  &  
55291.377 & 0.112799451(20) & \phantom{1}99.1 \\
{\it XMM} & EPIC-pn/SW &0650220201 & 2010 May 02 & 28.0/19.6  &  
55318.782 & 0.112799390(41) & \phantom{1}35.5 \\
\chandra\ & ACIS-S3/CC &     12108 & 2010 Aug 16 & 34.0/34.0  &  
55424.625 & 0.112799470(21) & \phantom{1}92.3 \\
{\it XMM} & EPIC-pn/SW &0650220901 & 2010 Oct 15 & 23.5/16.4  &  
55484.109 & 0.112799519(44) & \phantom{1}47.0 \\
{\it XMM} & EPIC-pn/SW &0650221001 & 2010 Oct 15 & 23.5/16.4  &  
55484.987 & 0.112799462(39) & \phantom{1}56.7 \\
{\it XMM} & EPIC-pn/SW &0650221101 & 2010 Oct 19 & 26.5/18.6  &  
55488.332 & 0.112799518(40) & \phantom{1}50.5 \\
{\it XMM} & EPIC-pn/SW &0650221201 & 2010 Oct 25 & 24.5/17.2  &  
55494.228 & 0.112799486(35) & \phantom{1}62.1 \\
{\it XMM} & EPIC-pn/SW &0650221301 & 2010 Nov 12 & 23.5/16.5  &  
55512.524 & 0.112799391(52) & \phantom{1}44.7 \\
{\it XMM} & EPIC-pn/SW &0650221401 & 2010 Dec 20 & 27.2/19.0  &  
55550.159 & 0.112799450(35) & \phantom{1}63.3 \\
\chandra\ & ACIS-S3/CC &     12109 & 2011 Feb 04 & 33.0/33.0  &  
55596.837 & 0.112799445(27) & \phantom{1}73.4 \\
{\it XMM} & EPIC-pn/SW &0650221501 & 2011 Apr 12 & 30.0/21.0  &  
55663.857 & 0.112799449(30) & \phantom{1}56.6 \\
{\it XMM} & EPIC-pn/SW &0657600101 & 2011 May 18 & 36.5/25.6  &  
55699.925 & 0.112799480(17) & \phantom{1}95.6 \\
\chandra\ & ACIS-S3/CC &     12541 & 2011 Aug 11 & 33.0/33.0  &  
55784.655 & 0.112799412(22) & \phantom{1}80.4 \\
{\it XMM} & EPIC-pn/SW &0657600201 & 2011 Nov 08 & 37.2/26.1  &  
55873.289 & 0.112799459(28) & \phantom{1}42.2 \\
\chandra\ &ACIS-S3/CC  &12542,14395& 2012 Feb 18,19& 33.1/33.1  &  
55975.446 & 0.112799481(35) &  \phantom{1}65.8 \\
{\it XMM} & EPIC-pn/SW &0657600301 & 2012 Apr 10 & 35.3/24.7  &  
56027.022 & 0.112799445(17) & \phantom{1}96.5
\enddata
\tablenotetext{a}{Barycentric period derived from a $Z^2_1$ test.
The \citet{lea83} uncertainty on the last digits is in parentheses.}
\label{timinglog}
\end{deluxetable*}

The \chandra\ observations used the Advanced Camera for Imaging and
Spectroscopy (ACIS-S3) in continuous-clocking (CC) mode to provide
time resolution of 2.85~ms. All data were reprocessed from the level~1
event files with the coordinates corrected for the proper motion of
\puppsr\ given in Table~\ref{ephemeris}, and analyzed using the latest
calibration files and software (CIAO 4.4/CALDB 4.4.8).  Reprocessing
with a source position that is accurate to $< 1$ pixel
($<0.\!^{\prime\prime}5$) ensures that the time assignment is precise
to $\lesssim 3$~ms.  All of the \xmm\ observations used the pn
detector of the European Photon Imaging Camera (EPIC-pn) in ``small
window'' (SW) mode to achieve 5.7~ms time resolution, and an absolute
uncertainty of $\approx 3$~ms on the arrival time of any photon. 
Each data set was examined and cleaned of intervals of high particle
background due to solar activity, as necessary. Two pairs of data sets
that were acquired on consecutive days were merged to improve their
statistics. The photon arrival times from all data were transformed to
Barycentric Dynamical Time (TDB) using the coordinates of the pulsar
corrected for proper motion.

For \puppsr, the pulsed signal strength is a strong function of
energy, not only because of its spectrum relative to the background
but because of the cancellation by emission from the opposite pole, as
described in \cite{got10}. For each observation listed in
Table~\ref{timinglog} we extracted source photons using an aperture
centered on the source and optimized for the signal strength in the
hard $1.5-4.5$~keV energy band. For the \xmm\ observations we used an
aperture of radius of $30^{\prime\prime}$.  For the \chandra\ CC-mode
observations we selected five columns ($2.\!^{\prime\prime}4$). We
also examined the soft, phase shifted $0.5-1.0$~keV band. However,
these data are noisier and their use did not in the end improve the
timing results significantly.  The phase cancellation effect also
prevents pulsations from being detected by the \chandra\ HRC, which
has insufficient energy resolution.

As in our previous timing studies of CCOs \citep{hal10a,hal11a}, we
employed two complementary approaches to fitting an ephemeris.  First,
we used the $Z^2_1$ (Rayleigh) test \citep{str80,buc83} in a coherent
analysis of the entire set of 17 observations.  Beginning with the
closely spaced set spanning 2010 October 15--19, the $Z^2_1$ test
determined the pulse frequency with sufficient accuracy to connect in
phase uniquely to the next observation.  This procedure was iterated
by adding each subsequent observation, and including a frequency
derivative when it became evident.  We also worked backward in time,
incorporating all 17 observations in the resulting unique ephemeris.
The fitted frequency derivative is $\dot f = (-6.94 \pm 0.28) \times
10^{-16}$ Hz~s$^{-1}$, where the $1\sigma$ uncertainty comes from the
$\Delta Z^2_1 = -2.3$ contour around the peak power in ($f,\dot f$)
space.

The second method also started with the $Z^2_1$ test statistic, this
time to find the period and pulse profile separately at each epoch.
The 17 profiles were cross-correlated, shifted, and summed to create a
master pulse profile template.  The process was iterated to generate a
more accurate template and a set of time-of-arrival (TOA) measurements
and their uncertainties for each epoch.  These TOAs were fitted with a
quadratic model in frequency and frequency derivative using a $\chi^2$
fitting routine to minimize their phase residuals.  We searched for a
coherent phase-connected solution over a grid of $f$ and $\dot f$
covering the range $f = 8.8652906 \pm 0.0000016$~Hz (at epoch MJD
55,580) and $-3.1 \times 10^{-14} < \dot f< 1.9 \times 10^{-14}$
Hz~s$^{-1}$, with an oversampling factor of 10 for accuracy. This
range corresponds to the $3\sigma$ limits of an incoherent fit to all
of the measured frequencies, including the 2001 discovery
observations.  The resulting frequency derivative from TOA fitting,
$\dot f = (-7.29 \pm 0.28) \times 10^{-16}$ Hz~s$^{-1}$, is consistent
with the value found above from the coherent $Z^2_1$ search.  We adopt
the TOA result for the final timing solution listed in
Table~\ref{ephemeris}.

\begin{figure}
\centerline{
\hfill
\includegraphics[width=0.9\linewidth,clip=]{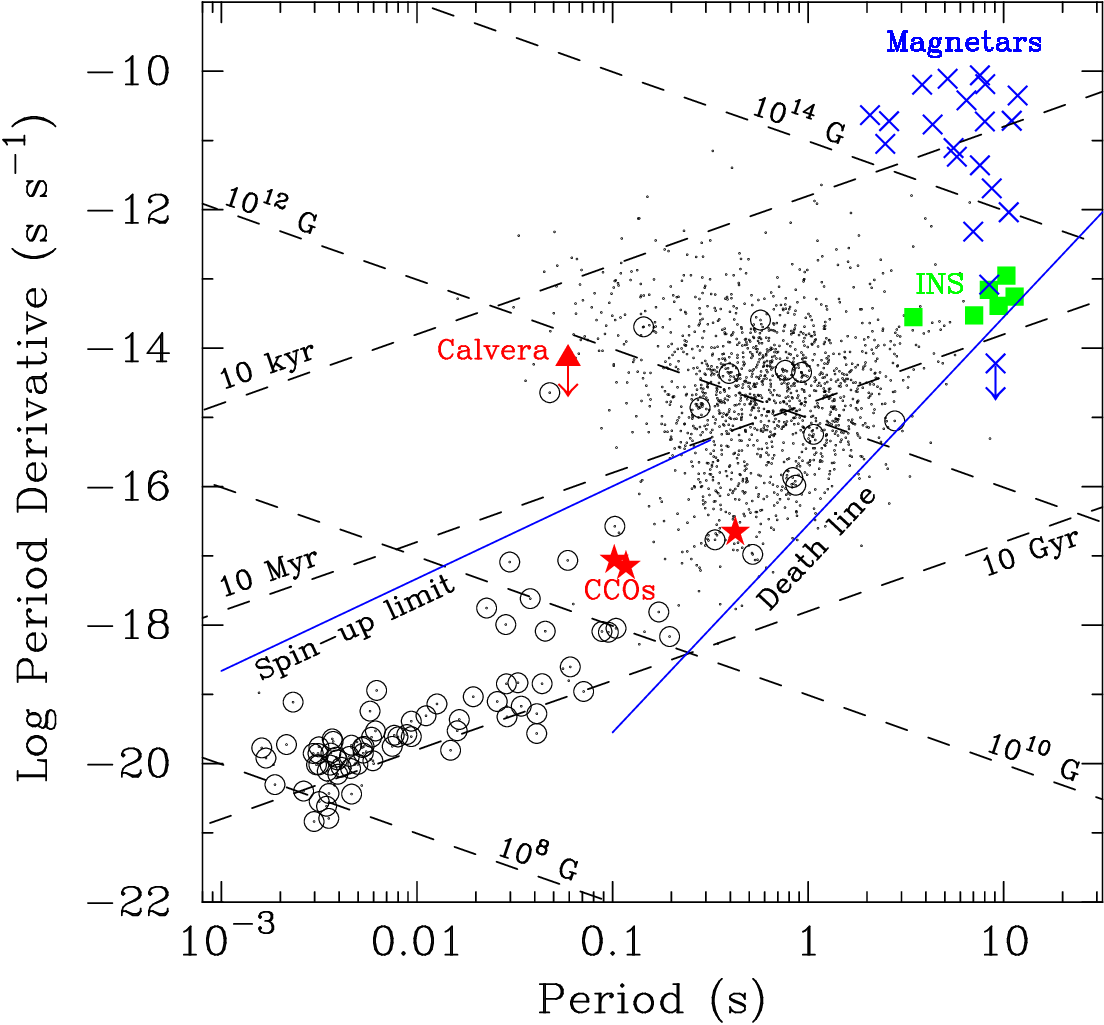}
\hfill
}
\caption{
$P-\dot P$ diagram of isolated pulsars (dots), binary radio pulsars
(circled dots), and other types of isolated X-ray pulsars
(colored symbols).
The CCO pulsars (red stars) in \snr\ and \psnr\
have virtually the same spin parameters.
The upper limit on Calvera's $\dot P$ is from \citet{hal11b}.
The radio pulsar death line $B/P^2 = 1.7 \times 10^{11}$ G~s$^{-2}$ of
\citet{bha92} is indicated.
The \citet{van87} spin-up limit for recycled pulsars corresponds to
$P({\rm ms}) = 1.9\,(B/10^9\,{\rm G})^{6/7}$.
The exponent in this equation corrects a typographical error 
in the caption to Figure~7 of \citet{hal10a},
although the corresponding line in the Figure was correct.
}
\label{ppdot}
\end{figure}

The observed $\dot P = (9.28\pm 0.36)\times 10^{-18}$ can 
now be split into the sum of its intrinsic and kinematic
contributions, $\dot P = \dot P_{\rm int} + \dot P_k$.
Since we determined in Section~2 that
$\dot P_k = (2.24 \pm 0.72)\times 10^{-18}$,
the intrinsic period derivative is
$\dot P_{\rm int} = (7.04 \pm 0.80)\times 10^{-18}$.
Parenthetically, we note that the small observed period derivative
is independent evidence that the proper motion of the pulsar is
not as large as the value originally quoted by \cite{win07},
$\mu = 165\pm 25$~mas~yr$^{-1}$.  If so, and if $d=2.2$~kpc, 
$\dot P_ k$ would be $(1.64 \pm 0.55) \times 10^{-17}$,
requiring $\dot P_{\rm int}$ to be negative, i.e., the pulsar
would be spinning up.

In the vacuum dipole spin-down formalism,
the values of $P$ and $\dot P_{\rm int}$
imply a surface magnetic field strength 
$B_s = 3.2 \times 10^{19}(P\dot P)^{1/2}~G = 2.9 \times 10^{10}$~G,
a spin-down luminosity $\dot E = -I\Omega\dot\Omega =
4\pi^2I\dot P/P^3 = 1.9 \times 10^{32}$~erg~s$^{-1}$,
and characteristic age $\tau_c \equiv P/2\dot P = 254$~Myr.
\puppsr\ is nearly identical in its spin properties to
\psr\ in \snr, as shown in Figure~\ref{ppdot}.
The uncertainties in distance and proper motion
have only a small effect on the
derived magnetic field.  For a smaller distance of 1~kpc, 
$\dot P_ k$ is reduced to $1.02\times 10^{-17}$,
and $B_s = 3.1 \times 10^{10}$~G.  An absolute upper limit
regardless of distance and proper motion is
$B_s < 3.3 \times 10^{10}$~G.

\begin{figure}
\centerline{
\hfill
\includegraphics[angle=270,width=0.9\linewidth,clip=]{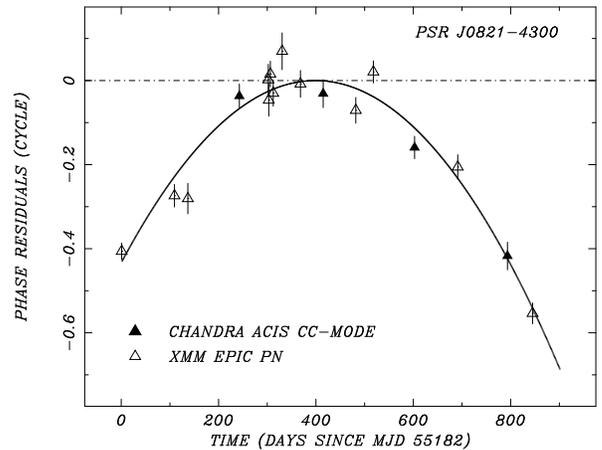}
\hfill
}
\caption{For the timing observations of \puppsr\ listed in
Table~\ref{timinglog}, pulse-phase residuals from the linear term
(dash-dot line) of the phase ephemeris presented in Table~\ref{ephemeris}.
The quadratic term (solid line) contributes $\approx \pm 0.5$ cycles to
the ephemeris over the 2.3 years of timing.}
\label{residuals}
\end{figure}

The phase residuals from the best-fit solution are shown in
Figure~\ref{residuals}.  The weighted rms of the phase residuals is
5.1~ms, or 0.045 pulse cycles, which is comparable to the individual
measurement errors (average $\sigma = 3.6$~ms).  It is not clear
if there is any real timing noise and/or systematic errors in the TOAs.

\begin{figure}
\centerline{
\hfill
\includegraphics[angle=270,width=0.9\linewidth,clip=]{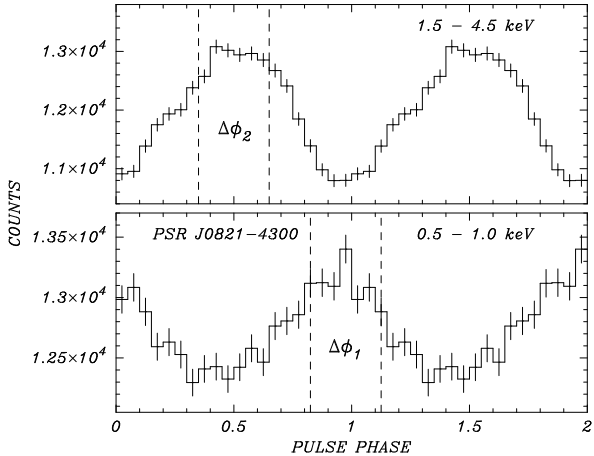}
\hfill
}
\caption{Summed pulse profiles of \puppsr\ in the $1.5-4.5$~keV band
(top) and the $0.5-1.0$~keV band (bottom) using all of the observations
listed in Table~\ref{timinglog}, folded according to the ephemeris of
Table~\ref{ephemeris}.  These hard and soft pulse profiles are out of
phase by $\phi = 0.45 \pm 0.02$ cycles. The intervals between the 
vertical lines ($\Delta\phi=0.3$~cycles) correspond to the two phase 
regions used in the phase resolved analysis of Section~4.3 and Table~\ref{phasetable}.}
\vspace{0.1in}
\label{lightcurve}
\end{figure}

The light curves of \puppsr\ in the soft and hard bands are shown in
Figure~\ref{lightcurve}.  These were derived by folding all the timing
data on the best fitting ephemeris given in Table~\ref{ephemeris}.  As
revealed by a cross-correlation, the soft and hard pulses are out of
phase by $0.45 \pm 0.02$ cycles, consistent with that found by
\cite{del12} using the 2009 December and 2010 April \xmm\ data. 

The energy dependence of the pulsar modulation provided an important
diagnostic for modeling the viewing geometry and surface emission of
\puppsr\ \citep{got10}. By analyzing the lightcurve in narrower energy
bands than in Figure~\ref{lightcurve}, we can resolve the signal
modulation and phase for \puppsr\ over the $0.3-5$~keV range.  The
data were grouped into 23 energy bands that are at least $100$~eV in
width and have a signal-to-noise $N_s/\sqrt{N_s+N_b} > 100$, where
$N_s,N_b$ are the source and background counts, respectively.  We used
$Z_1^2$ to provide a model of the unbinned lightcurve.  The first
Fourier component is a reasonable estimate as the lightcurve is
sinusoidal in each energy band to within counting statistics.  The
error bar for the phase is calculated by cross-correlation, with the
profile of Figure~\ref{lightcurve} serving as a template.

\begin{figure}
\begin{center}
\includegraphics[angle=270,width=0.8\linewidth]{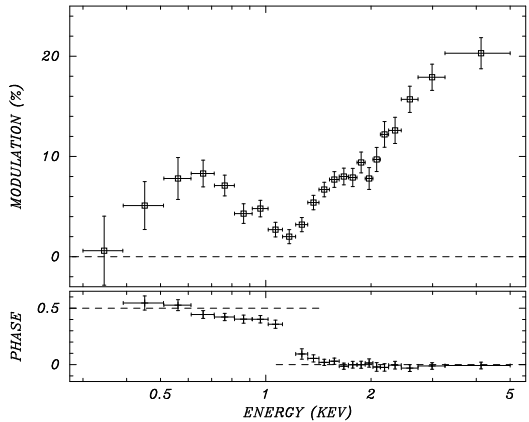}
\end{center}
\caption{
Modulation and pulse phase as a function of energy for \puppsr\ using
all of the observations listed in Table~\ref{timinglog}, folded
according to the ephemeris of Table~\ref{ephemeris}. The modulation
(top) and pulse phase (bottom) reproduce the form predicted for the
antipodal model of \cite{got10}.  Two undefined phase points are omitted.
}
\label{modulation}
\end{figure}

The result is presented in Figure~\ref{modulation}. As the energy
dependent modulation decreases, the phase becomes undefined in two
energy bands; these two phase points are not plotted. The modulation
is qualitatively similar to that predicted by the antipodal model of
\cite{got10} (cf. their Figure 6), providing confirmation of the basic
model.  The prediction was based on fitting the modulation in only
three energy bands, using much less data, and differs somewhat from
the new, resolved data having an order-of-magnitude more counts.  In
particular, the energy of the minimum modulation is lower (1.12 vs.
1.28 keV), and the form of the modulation is more complex than
predicted.  Furthermore, the phase is seen to drift at lower energies
and the transition is not as sharp compared to the antipodal case, in
which the phase was statistically either 0.0 or 0.5. The observed
characteristics likely imply that the hotspots are offset from a
strictly antipodal geometry.
The high quality data presented
herein should allow a far more detailed modeling of the surface
emission of \puppsr.

\section{Spectral Analysis}

Previously, we modelled the original (year 2001) \xmm\ spectra of
\puppsr\ as surface blackbody emission from two antipodal spots of
different temperatures and areas.  Crucially, this model is also able
to account for the observed energy-dependent pulse modulation and
phase shift \citep{got09,got10}.  Additionally, we found evidence of a
spectral line feature around 0.8~keV, which more recent data obtained
by \cite{del12} suggests has a time-variable centroid energy.  With
the increased quantity of data now in hand on \puppsr, we can
re-examine this spectral feature, first by combining all 13 \xmm\
observations presented in Table~\ref{timinglog} plus the two
observations obtained in 2001, and then by testing for any variability
among the observations.

\subsection{Summed \xmm\ Spectrum}

The spectral analysis presented here uses exclusively the data
collected with the EPIC pn.  Although background is much reduced in
the EPIC MOS, this instrument is less sensitive at soft energies,
collecting 4.7 times fewer photons than the EPIC pn in the $0.3-1$~keV
band.  Furthermore, much of the EPIC MOS data were lost as the high
surface brightness of the \psnr\ SNR frequently triggered an automatic
shutoff of that detector.  We also neglect the \chandra\ ACIS spectra
here because of their even poorer low-energy sensitivity and the
increased background and other uncertainties involved in the analyzing
the \chandra\ CC-mode data.  Lastly, one ACIS image taken in timed
exposure mode (ObsID 750) suffers from pileup, and was not used.

For the EPIC pn data, the main technical issue is that the SNR background
exceeds the point source counts below 1~keV.  Surprisingly, the background
intensity and spectral shape are both strong functions of spacecraft
roll angle.  This effect is made evident because the \xmm\ data sets were
acquired in two narrow ranges of roll angle roughly 180$^{\circ}$ apart
associated with their respective visibility windows, with the time divided
nearly equally between the two.  We checked all of our spectral results
carefully for features that might be dependent on roll angle due to systematic
errors in background subtraction.  No such systematic effect were found,
which indicates that the background subtraction is reliable in general.

\begin{figure}
\centerline{
\includegraphics[angle=270,width=0.9\linewidth,clip=]{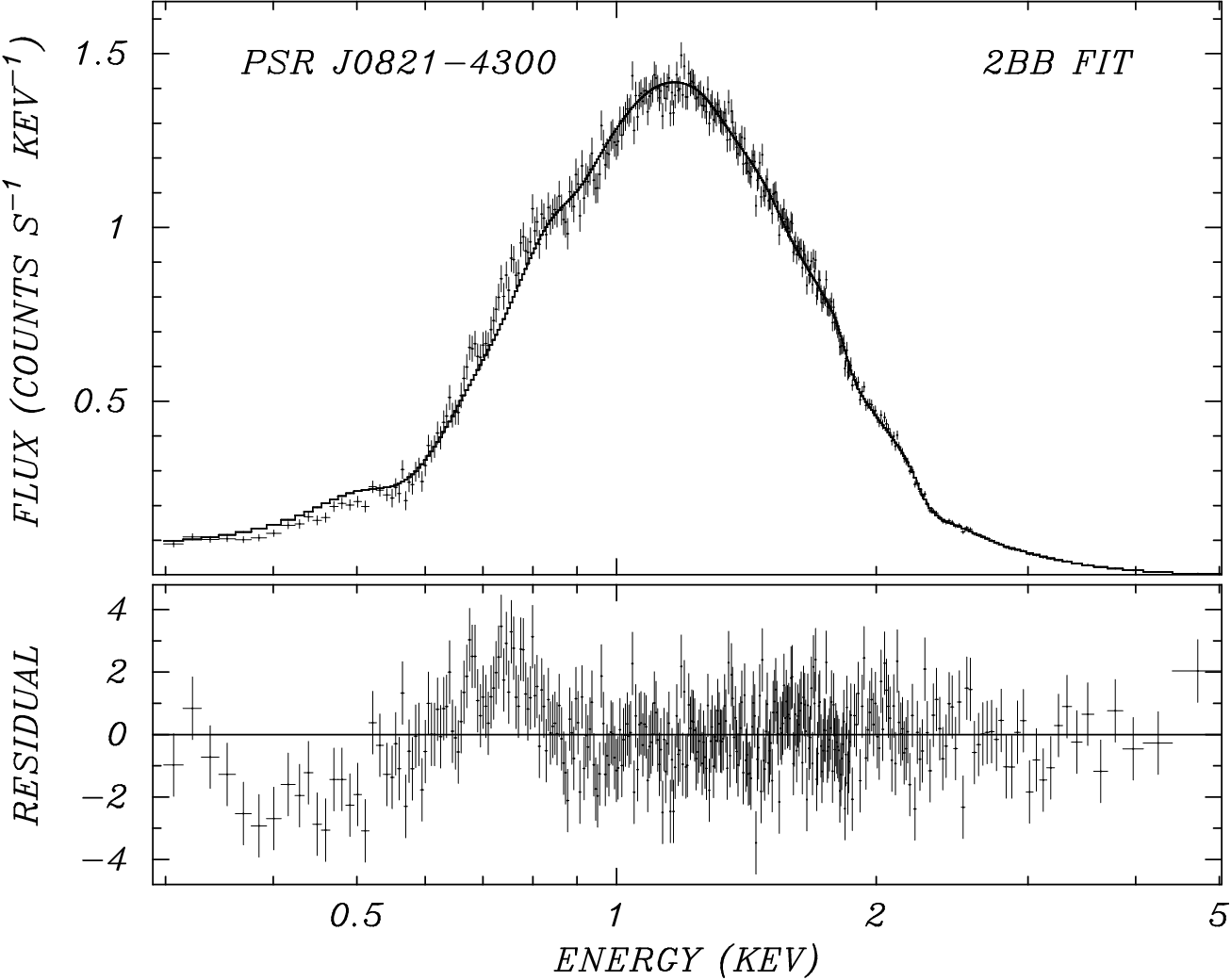}
}
\vspace{0.2in}
\centerline{
\includegraphics[angle=270,width=0.9\linewidth,clip=]{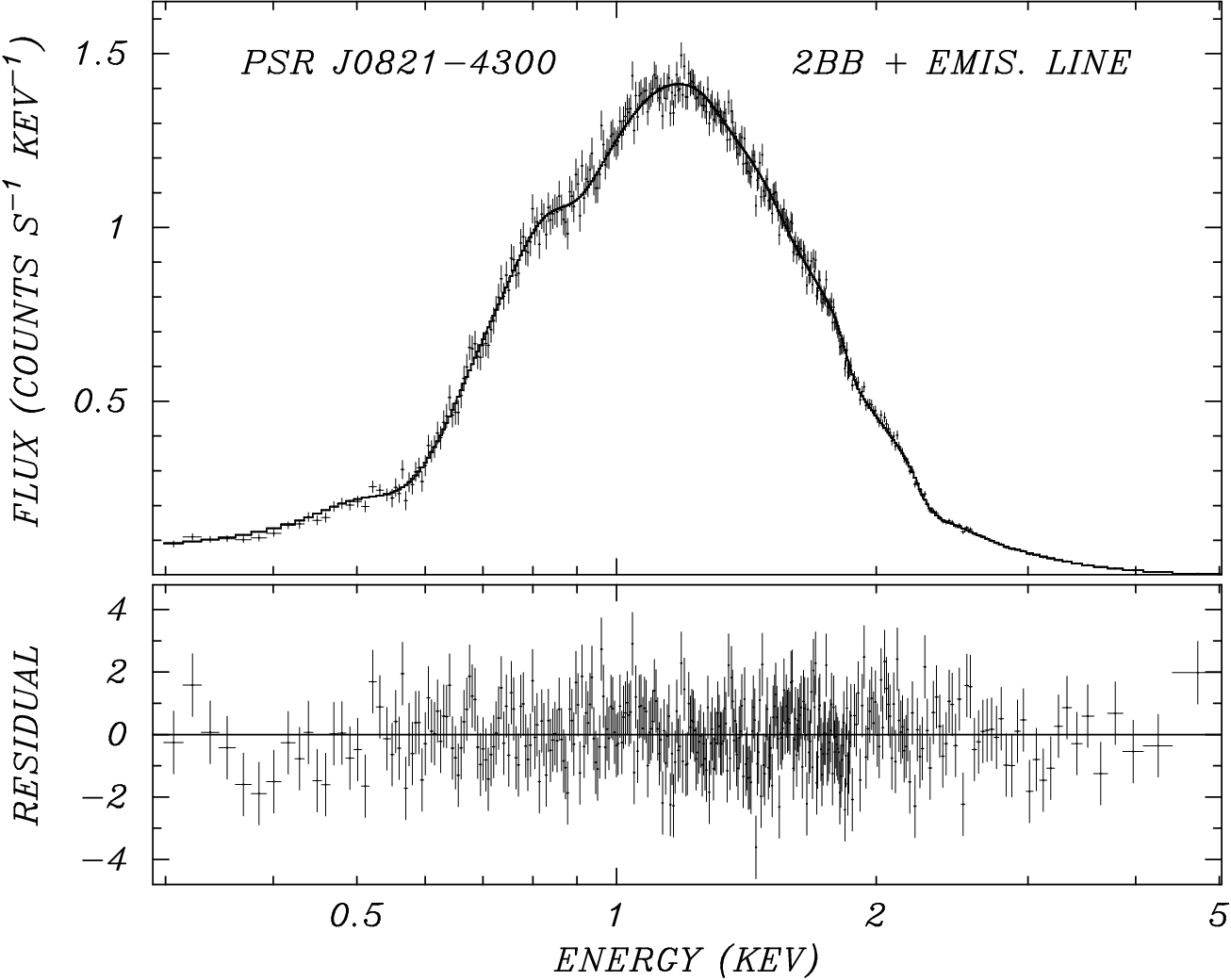}
}
\caption{
(a) EPIC pn spectrum of the 16 summed \xmm\
observations of \puppsr\ fitted to a double blackbody model.
The residuals from the fit are shown in the bottom panel. 
(b) The same spectrum fitted with a double blackbody model 
plus Gaussian emission line. The parameters of these fits
are given in Table~\ref{speclines}.
}
\label{spectrum}
\end{figure}

We extracted spectra for each observation from the EPIC pn detector
using an aperture of radius $0\farcm3$ and a concentric background
annulus of $0\farcm5<r<0\farcm6$, selecting only events with ${\tt
  PATTERN} \le 4$ and FLAG$=0$.  The data were filtered to exclude time
intervals of high background identified by count rate $>0.1$~s$^{-1}$
in the $10-12$ keV energy band.  An inspection of the pattern
distribution of single and double events shows no evidence of pile-up
and suggests that a lower energy bound in the range $0.3-0.4$~keV is
acceptable.  Response matrices and effective area files were generated
for each observation using the SAS software suite. We combined data
from all observations using the FTOOL {\it addascaspec} to produce a
single source spectrum and associated files.  The combined spectrum
was grouped to include at least 1000 counts per channel and was fitted
using {\tt XSPEC} v12.21 to a two blackbody model with interstellar
absorption over the energy range $0.3-5$~keV (see
Figure~\ref{spectrum} and Table~\ref{speclines}).

\begin{deluxetable*}{lccccccc}
\tabletypesize{\scriptsize}
\tablewidth{0pt}
\tablecaption{Models for the Summed \xmm\ Spectrum of \puppsr}
\tablehead{
     Model                        &     Two Blackbody &           $+$ Emis. line&       $+$ Abs. line&  \multicolumn{2}{c}{$+$~Two~Abs.~lines}  & \multicolumn{2}{c}{$+$~Cyclabs}
} 
\startdata                       
 $N{\rm_H}$ ($10^{21}$~cm$^{-2}$) &   $3.8\pm0.1$   & $4.3\pm0.3$   & $3.2\pm0.2$   & $2.9\pm0.4$   & \dots      & $2.8\pm1.01$  & \dots        \\
 $k{\rm T}_w$ (keV)               &   $0.26\pm0.01$ & $0.25\pm0.01$ & $0.29\pm0.01$ & $0.29\pm0.01$ & \dots      & $0.28\pm0.03$ & \dots        \\
 $k{\rm T}_h$ (keV)               &   $0.46\pm0.01$ & $0.45\pm0.01$ & $0.49\pm0.02$ & $0.49\pm0.03$ & \dots      & $0.47\pm0.02$ & \dots        \\
 $L_w({\rm bol})$ ($10^{33}$~erg~s$^{-1}$)\tablenotemark{a}&   $3.3\pm0.1$   & $3.6\pm0.2$   & $3.1\pm0.2$   & $3.0\pm0.2$   & \dots      & $3.2\pm0.8$   & \dots        \\
 $L_h({\rm bol})$ ($10^{33}$~erg~s$^{-1}$)\tablenotemark{a}&   $2.0\pm0.2$   & $2.0\pm0.2$   & $1.4\pm0.3$   & $1.4\pm0.3$   & \dots      & $1.7\pm0.4$   & \dots        \\
 $A_w$ (km$^{2}$)                 &   $72\pm11$     & $89\pm18$     & $44\pm6$      &	$40\pm8$    & \dots      & $53\pm14$     & \dots        \\
 $A_h$ (km$^{2}$)                 &   $4.4\pm0.9$   & $4.5\pm0.9$   & $2.5\pm0.8$   & $2.4\pm1.0$   & \dots      & $3.4\pm0.8$  & \dots        \\
 $E_0$ (keV)                      &     \dots       & $0.75\pm0.01$ & $0.46\pm0.05$ & $0.46\pm0.01$ & $2E_o$     & $0.46\pm0.01$   &  $2E_o$      \\ 
 Width\tablenotemark{b} (eV)      &     \dots       & $75\pm20$     & $85\pm50$     &	$106\pm20$  &	$34-62$  & $53-97$       &  $35-290$    \\
 EW (eV)                          &     \dots       & $53\pm10$     &    \dots      &	\dots       & \dots      & \dots	 & \dots        \\
 $\tau_o$\tablenotemark{c}        &     \dots       & \dots         & $0.1-0.9$     &	$0.6-1.3$   &	$<0.035$ & $0.9-1.7$     & $<0.14$  \\
     $\chi^2({\rm DoF})$          &  $1.50(359)$    &  1.08(356)    & $1.16(356)$   & \multicolumn{2}{c}{$1.08(354)$}&   \multicolumn{2}{c}{$1.12(354)$} 
\enddata								 
\tablecomments{The $1\sigma$ uncertainties for three interesting parameters
($\Delta \chi^2 = 3.53$) are given.}
\tablenotetext{a}{Blackbody bolometric luminosity for a distance of 2.2 kpc.}
\tablenotetext{b}{Gaussian $\sigma$ for the emission or absorption lines,
natural width $W$ for the cyclotron absorption model \citep{mak90,mih90}.}
\tablenotetext{c}{Optical depth at line center.}
\label{speclines}
\end{deluxetable*}

The resulting high signal-to-noise ratio of the fitted spectrum reveals
significant features that are unaccounted for by the two blackbody model.
The best fit model, with $N_{\rm H} = (3.8\pm0.01)\times 10^{21}$~cm$^{-2}$,
$kT_w = 0.26\pm0.01$~keV, and $kT_h = 0.46\pm0.01$~keV, has reduced
$\chi^2_{\nu} = 1.50$ for 359 degrees of freedom (DoF), which is formally
unacceptable.  The deviations are evident in structure in the residuals
in Figure~\ref{spectrum}a.  Adding a Gaussian emission line to the
model as suggested by our previous work improves the fit
to $\chi^2_{\nu} = 1.08$ for 358 DoF (Figure~\ref{spectrum}b).
The centroid energy of the line is $0.75 \pm 0.01$~keV, and its
equivalent width is $53\pm10$~eV.

\begin{figure}
\centerline{
\includegraphics[angle=270,width=0.9\linewidth,clip=]{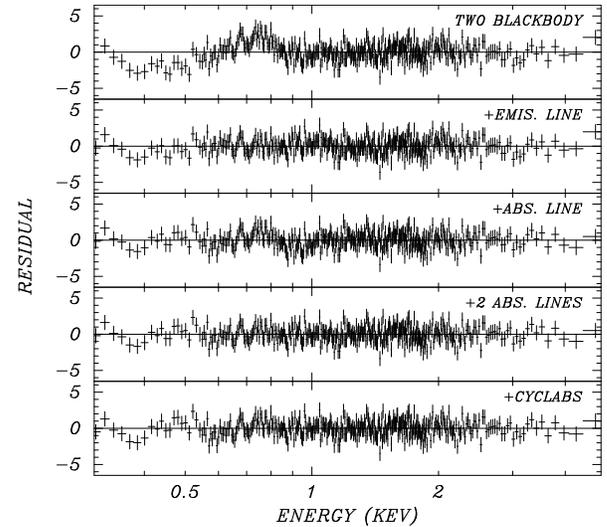}
}
\caption{
Same as Figure~\ref{spectrum} but showing the residuals from fits
to a double blackbody model plus, from top to bottom: no
line, a Gaussian emission line, a Gaussian absorption line,
two Gaussian absorption lines, and a cyclotron absorption line model.
The parameters of the fits are given in Table~\ref{speclines}.}
\label{spectrum2}
\end{figure}

Considering the shape of the residuals from the two blackbody fit in
Figure~\ref{spectrum}a, an alternative hypothesis is that an
absorption feature, or features, is responsible.  Accordingly we
applied a Gaussian absorption line, two Gaussian absorption lines, and
finally, the cyclotron absorption model of \citet{mak90} and
\citet{mih90}, which is available in {\tt XSPEC} as {\tt cyclabs}.  As
shown in Table~\ref{speclines} and Figure~\ref{spectrum2}, all of
these models gave acceptable fits.  Two absorption lines, when fitted
independently, are separated by nearly a factor of 2 in energy, which
is suggestive of the fundamental and first harmonic in a cyclotron
model.  Therefore we fixed the ratio of their centroid energies at 2,
with the result that their centroids are at 0.46~keV and 0.92~keV,
bracketing the energy previously ascribed to an emission line. The
{\tt cyclabs} model, which also fixes this ratio, yields the same
centroid energies.  We conclude that, from the phase-averaged spectra
alone, it is not possible to distinguish an emission feature from one
or two absorption lines, which leaves the physical interpretation
uncertain.

The fundamental energy of the electron cyclotron
resonance falls at
$$E_0 = 1.16\,(B/10^{11}\,{\rm G})/(1+z)\ {\rm keV},\eqno(2)$$ where
$z$ is the gravitational redshift.  Assuming a typical value of $z=0.3$,
and if $B \approx B_s = 2.9 \times 10^{10}$~G, the equatorial field
from the vacuum dipole spin-down result, then $E_0 \approx 0.26$~keV 
is expected.
When compared with $E_0 = 0.46$~keV from the absorption-line fit, or $0.75$~keV
from the emission-line fit, this hints that the local surface field where
the line is formed is larger than the equatorial dipole field, but is perhaps
more compatible with the field at the pole, $B_p = 2\,B_s$.
Other factors possibly affecting a comparison between the dipole
spin-down $B$-field and the surface $B$-field from the cyclotron energy will
be explored in Section~6.2.

The addition of emission or absorption lines to the model
near the low-energy end of the \xmm\ spectrum affects the
fitted value of the column density $N_{\rm H}$, with
values ranging from $(2.8-4.3) \times 10^{21}$~cm$^{-2}$
for the different models in Table~\ref{speclines}.
Therefore, it is possible that comparison with
independent measurements of $N_{\rm H}$ may suggest
a preference for one or another of these models. 
For example, the 21~cm \ion{H}{1} emission in the foreground
of \psnr\ amounts to $N_{\rm HI} = 2.5 \times 10^{21}$~cm$^{-2}$
according to \citet{rgj03}.  To obtain this value, they integrated
the \ion{H}{1} line emission in the radial velocity range
$-10$ to $+16$ km~s$^{-1}$, the latter velocity corresponding
to their assumed 2.2~kpc distance of \psnr.
Comparison with the values of $N_{\rm H}$ in Table~\ref{speclines} 
tends to favor the X-ray spectral models that include absorption lines.
While this agreement is encouraging, it is subject to the caveat
that the 21~cm column density is formally a lower limit,
assuming as it does that the line is optically thin.
Better support for a low column density
comes from the \xmm\ RGS spectra of several regions of the \psnr\
SNR fitted by \citet{kat12}.  These require X-ray
$N_{\rm H}$ in the range $(2.58-2.85) \times 10^{21}$~cm$^{-2}$,
which also agrees closely with the $N_{\rm H}$ from our
absorption-line models for the pulsar spectrum.
\begin{figure}
\centerline{
\hfill
\includegraphics[angle=270,width=0.9\linewidth,clip=]{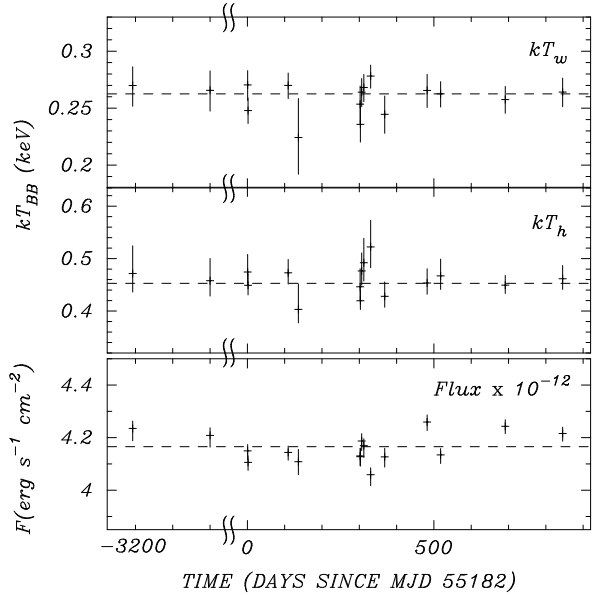}
\hfill
}
\caption{
Blackbody temperatures and flux in the $0.5-5$ keV band for the 16
individual \xmm\ observations of \puppsr, fitted to the two blackbody
model.  Data are taken from Table~\ref{spectable}.
Errors bars are 1$\sigma$.  The weighted mean values are indicated 
by the dashed lines.}
\label{fluxes}
\end{figure}

\begin{figure}
\centerline{
\includegraphics[angle=270,width=0.9\linewidth,clip=]{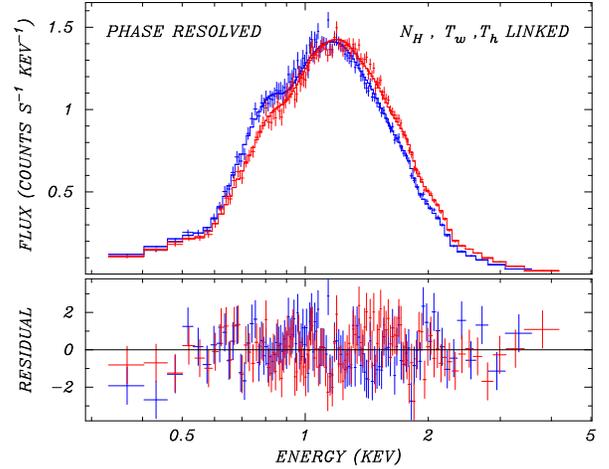}
}
\caption{Combined \xmm\ EPIC pn spectrum of \puppsr\ in two
pulse-phase intervals defined in Figure~\ref{lightcurve}
($\Delta\phi_1$, blue, $\Delta\phi_2$, red), fitted to a double
blackbody model plus Gaussian emission line.  The blackbody
temperatures and line centroid energy are linked between the two
spectra.  The fitted parameters are given in the last column of
Table~\ref{phasetable}.  The residuals to the fit are shown in the
bottom panel.}
\label{phasespecall}
\end{figure}

Finally, we note that the total luminosity and blackbody areas reported
in Table~\ref{speclines} differ from those presented in \citet{got09}, the
areas by a factor of two or more. This is a consequence of the best
fit values for each model fit. The derived blackbody areas depend on
blackbody temperature as $T^{-4}$, which itself is strongly
correlated with the fitted column density. Future work,
simultaneously modeling of the phase dependent spectra, should
better constrain the column density and temperatures, and
consequently provide a more accurate measurement of the blackbody
areas.

\subsection{Search for Variability}

To test for long-term variability of \puppsr, we also fitted
the individual \xmm\ spectra, searching for temperature variations
on the surface, for example.  The  spectra for each observation are
well-fitted by the two blackbody model, with parameters
listed in Table~\ref{spectable}; their statistics
do not warrant an additional line component.  For this comparison,
we held the absorbing column fixed at the value in 
Table~\ref{speclines} derived from the fit of the composite spectrum
to the two blackbody model. Figure~\ref{fluxes} displays the results.
No significant variation is evident in the flux or
in either of the two blackbody temperatures.

\begin{deluxetable*}{lcccccc}
\tabletypesize{\scriptsize}
\tablewidth{0pt}
\tablecaption{Individual \xmm\ Spectra of \puppsr}
\tablehead{
\colhead{Group}&\colhead{ObsID}&\colhead{Livetime} & 
\colhead{$kT_w$}&\colhead{$kT_h$}& \colhead{Flux ($10^{-12}$} & 
\colhead{$\chi^2_{\nu}({\rm DoF})$} \\
\colhead{}     & \colhead{}    & \colhead{(ks)}   & \colhead{(keV)} & 
\colhead{(keV)}& \colhead{erg s$^{-1}$ cm$^{-2}$)}  & \colhead{}
}
\startdata
1 &0113020101S & 15.2 & $0.27(0.25,0.29)$ & $0.47(0.44,0.52)$ &  $4.23(4.14,4.30)$ & $1.06(154)$\\
  &0113020301S & 16.1 & $0.27(0.25,0.28)$ & $0.46(0.43,0.50)$ &  $4.21(4.12,4.27)$ & $1.05(184)$\\
\hline
2 &0606280101S & 25.7 & $0.27(0.26,0.28)$ & $0.47(0.45,0.51)$ &  $4.15(4.09,4.20)$ & $1.19(282)$ \\
  &0606280101U & 24.5 & $0.25(0.24,0.26)$ & $0.45(0.43,0.47)$ &  $4.11(4.05,4.16)$ & $0.91(268)$\\
  &0606280201S & 24.6 & $0.27(0.26,0.28)$ & $0.47(0.45,0.50)$ &  $4.14(4.09,4.20)$ & $1.02(261)$ \\
  &0650220201S &  8.1 & $0.22(0.19,0.26)$ & $0.40(0.38,0.45)$ &  $4.11(4.01,4.20)$ & $0.88(94)$ \\
\hline
3 &0650220901S & 16.4 & $0.25(0.24,0.27)$ & $0.45(0.42,0.48)$ &  $4.13(4.06,4.19)$ & $1.01(191)$ \\
  &0650221001S & 15.5 & $0.24(0.22,0.25)$ & $0.42(0.40,0.44)$ &  $4.13(4.06,4.20)$ & $0.99(183)$ \\
  &0650221101S & 18.6 & $0.26(0.25,0.28)$ & $0.48(0.45,0.51)$ &  $4.19(4.12,4.25)$ & $0.97(222)$ \\
  &0650221201S & 17.1 & $0.27(0.26,0.28)$ & $0.49(0.46,0.54)$ &  $4.17(4.08,4.23)$ & $1.08(201)$ \\
  &0650221301S & 16.4 & $0.28(0.27,0.29)$ & $0.52(0.48,0.57)$ &  $4.06(3.98,4.12)$ & $1.21(182)$ \\
  &0650221401S & 15.5 & $0.24(0.23,0.26)$ & $0.43(0.41,0.46)$ &  $4.13(4.05,4.20)$ & $1.00(174)$ \\
\hline
4 &0650221501S & 20.3 & $0.27(0.25,0.28)$ & $0.45(0.43,0.48)$ &  $4.26(4.20,4.32)$ & $0.96(225)$ \\
  &0657600101S & 22.4 & $0.26(0.25,0.27)$ & $0.47(0.44,0.50)$ &  $4.13(4.07,4.19)$ & $0.99(242)$ \\
  &0657600201S & 26.1 & $0.26(0.25,0.27)$ & $0.45(0.43,0.47)$ &  $4.24(4.19,4.30)$ & $1.07(285)$ \\
  &0657600301S & 24.7 & $0.26(0.25,0.28)$ & $0.46(0.44,0.49)$ &  $4.22(4.16,4.27)$ & $0.94(262)$
\enddata
\tablecomments{Results for a fit to an absorbed, two-blackbody 
model with the column density held fixed at
$N_{\rm H} = 3.75 \times 10^{21}$~cm$^{-2}$, the value
obtained for the summed spectrum in Table~\ref{speclines}. The absorbed 
flux in the $0.5-5$~keV range is tabulated. The range of uncertainty
of each value is the $1\sigma$ confidence interval for two
interesting parameters.}
\label{spectable}
\end{deluxetable*}

We next test for pulse-phase dependence of the spectral line
feature(s) by generating spectra in two phase intervals, each of
$\Delta \phi=0.3$ cycles in width centered on the peaks of the soft
and hard light curves, respectively (see Figure~\ref{lightcurve}.)
This is motivated by the original finding in \citet{got09} that an
emission line is more strongly associated with the warm region, i.e.,
the soft phase of the pulse.  For simplicity, given the reduced counts
in the phased spectra, we used only the Gaussian emission-line model
as a representative for all of the possible line models.  The two
phase-resolved spectra combined from all epochs are shown in
Figure~\ref{phasespecall}, where they are fitted to the two blackbody
plus Gaussian emission-line model.  In the simultaneous fit, the two
temperatures and the line centroid energy are linked.  As shown in
Table~\ref{phasetable}, the equivalent width of the Gaussian line is
indeed about factor of 2 larger in the soft-phase spectrum than in the
hard phase.

\begin{figure*}
\centerline{
\hfill
\includegraphics[scale=0.35,angle=270]{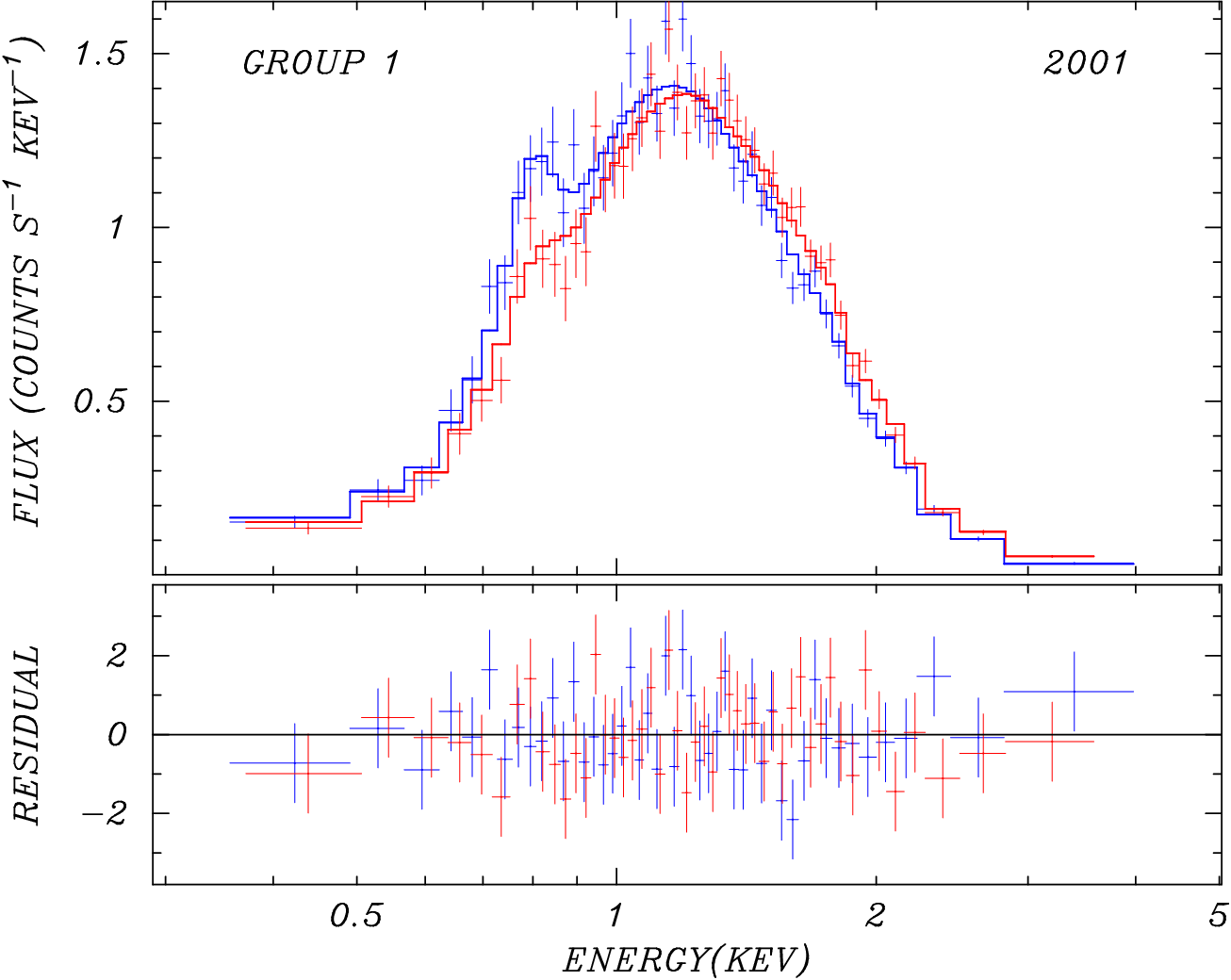}
\hfill
\includegraphics[scale=0.35,angle=270]{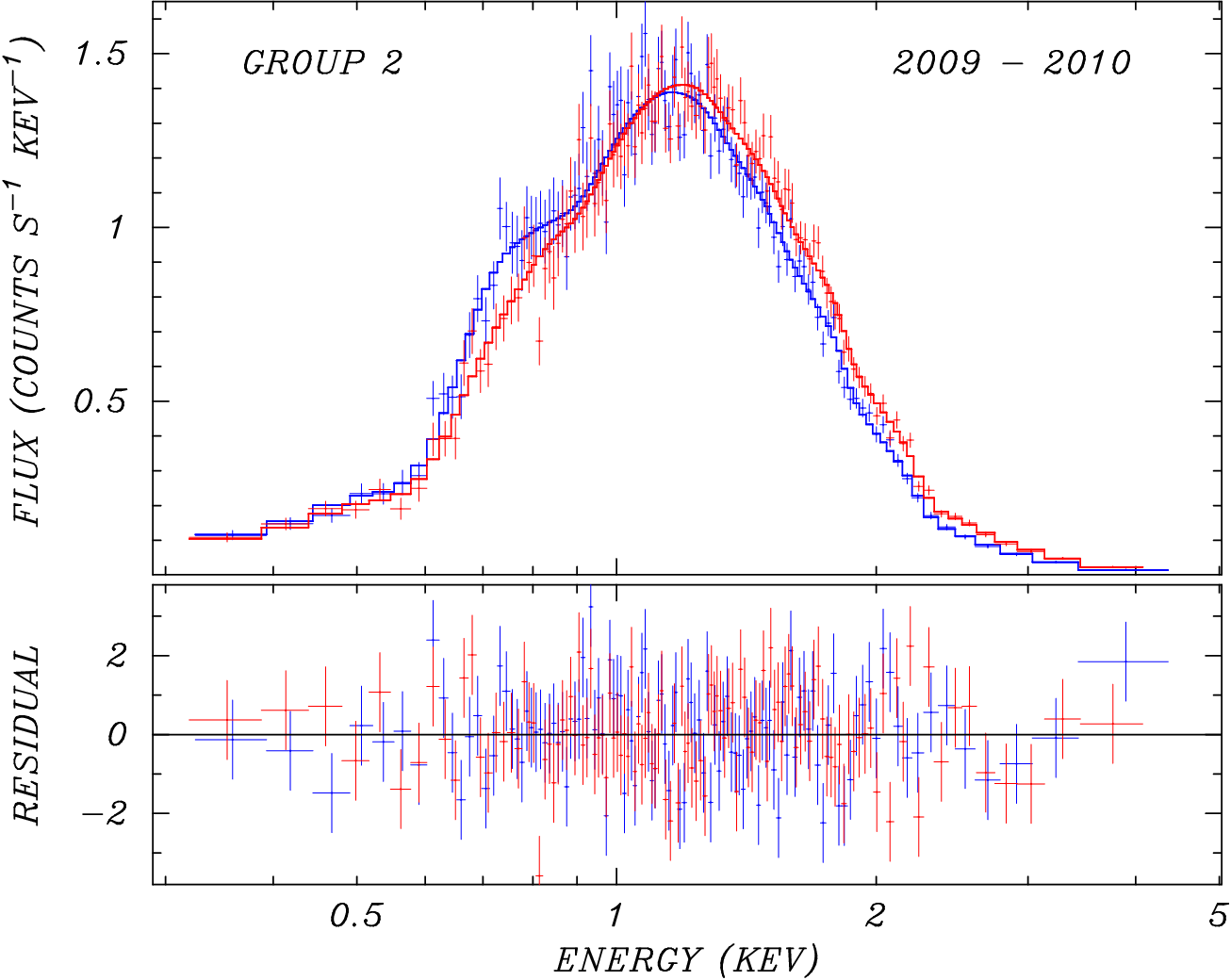}
\hfill
}
\vspace{0.2in}
\centerline{
\hfill
\includegraphics[scale=0.35,angle=270]{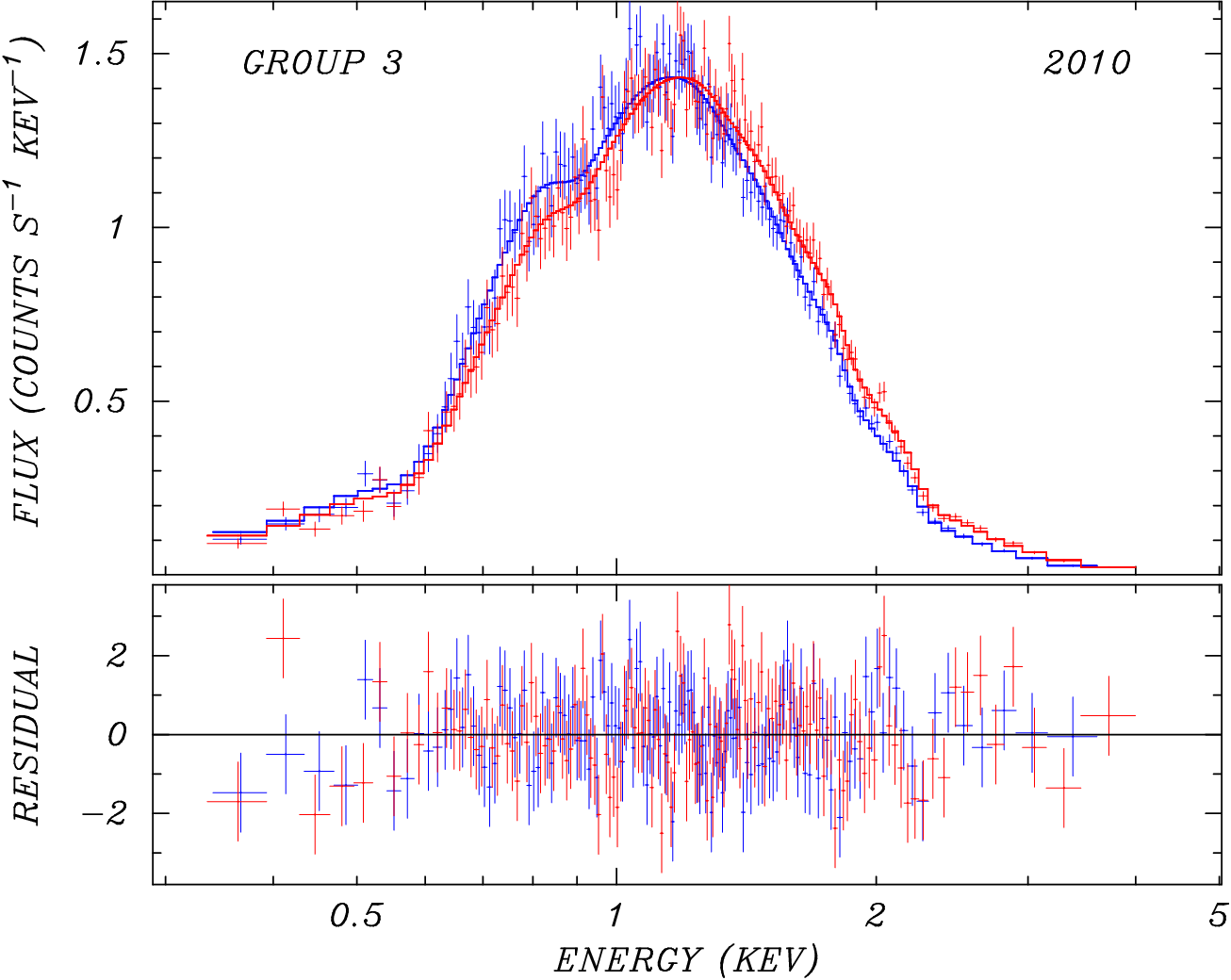}
\hfill
\includegraphics[scale=0.35,angle=270]{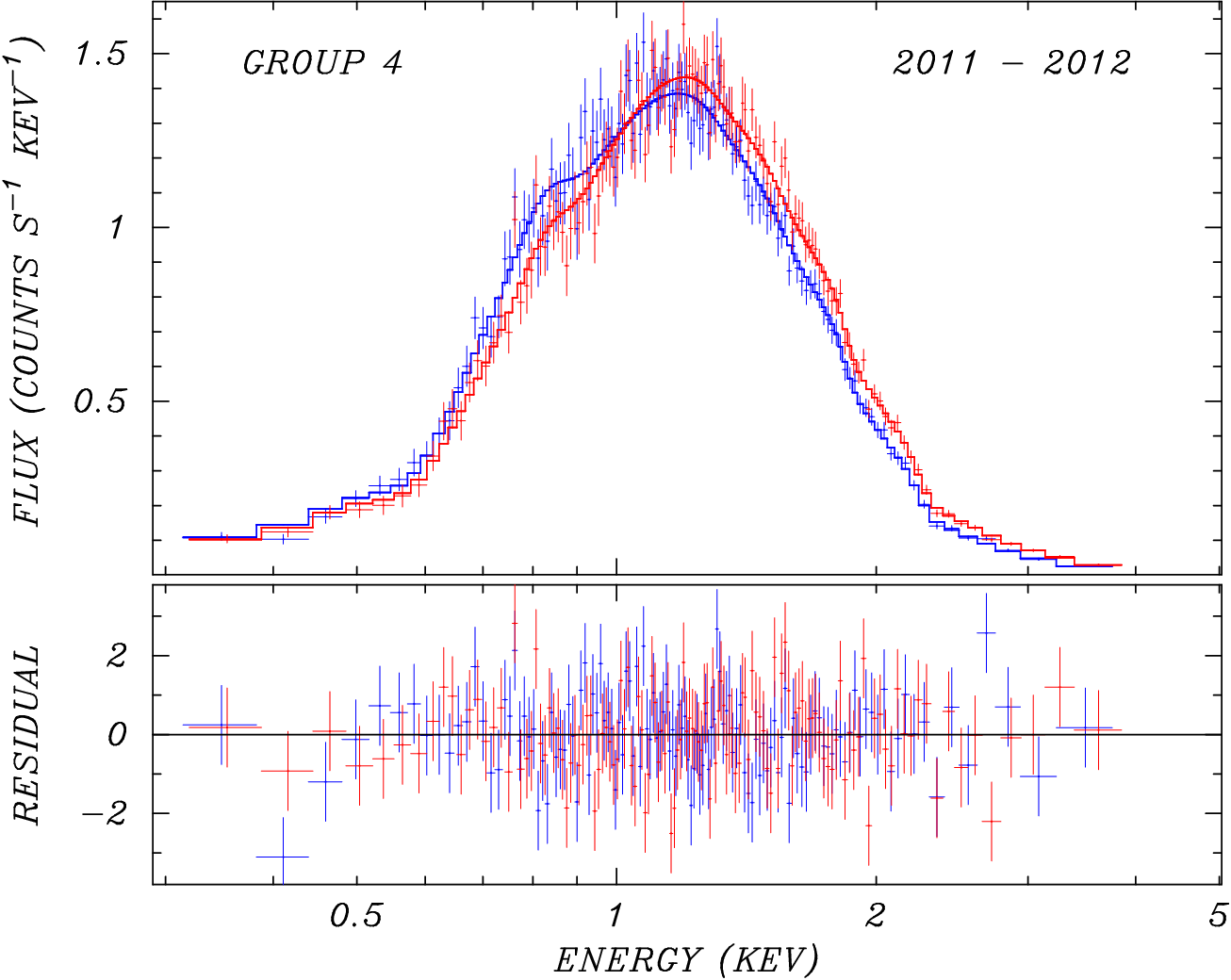}
\hfill
}
\caption{
\xmm\ EPIC pn spectra of \puppsr, grouped into the four 
sets as listed in Table~\ref{spectable},
and fitted to a double blackbody model with Gaussian emission line.
As in Figure~\ref{phasespecall}, red and blue represent the two
pulse-phase intervals defined in Figure~\ref{lightcurve}.
The residuals from the fits are shown in the bottom panels. 
The fitted parameters are given in Table~\ref{phasetable}.  
}
\label{phasespec}
\end{figure*}

\citet{del12} presented evidence that the emission-line centroid
had decreased from 0.80~keV to 0.73~keV between 2001 and 2009.
Here we extend the examination for variability of the spectral
feature, including its phase dependence,
by combining the \xmm\ spectra into four groups of adjacent observations
as indicated Table~\ref{spectable}.   We fitted these four sets
of spectra to the two blackbody plus Gaussian emission-line model,
with their temperatures and line width (sigma) linked between the two
phase intervals.
The four fits are displayed in Figure~\ref{phasespec},
and the fitted parameters for each set and their sum are
presented in Table~\ref{phasetable}.  It is clear that the 
temperatures and their contributed fluxes are steady in time, 
consistent with their phase-averaged behavior and with no time
dependence in their phase ratios.  On the other hand, the equivalent
width of the fitted Gaussian emission line shows evidence for having
increased between 2001 and 2009, while the line centroid energy decreased
from $0.79^{+0.02}_{-0.03}$~keV to $0.71^{+0.02}_{-0.03}$~keV.  While these
results are consistent with the \citet{del12} analysis of the
same data, there is little if any additional variability between
2009 and 2012.

\begin{deluxetable*}{lccccc}
\tabletypesize{\scriptsize}
\tablewidth{0pt}
\tablecaption{\xmm\ Phase-Resolved Spectra of \puppsr }
\tablehead{
\colhead{Model} & \colhead{Group 1} & \colhead{Group 2} & \colhead{Group 
3} & \colhead{Group 4} & \colhead{Sum}\\
\colhead{Parameter} & \colhead{2001 Apr--Nov} & \colhead{2009 Dec -- 2010 May}
& \colhead{2010 Oct--Dec} & \colhead{2011 Apr -- 2012 Apr} & 
\colhead{2001 Apr -- 2012 Apr}
}
\startdata
$kT_w$ (keV)                     &$0.23(0.21,0.25)$& $0.25(0.24,0.26)$& $0.24(0.23,0.25)$& $0.27(0.25,0.28)$& $0.25(0.24,0.25)$  \\
$kT_h$ (keV)                     &$0.41(0.39,0.44)$& $0.45(0.43,0.47)$& $0.44(0.42,0.46)$& $0.47(0.44,0.52)$& $0.44(0.43,0.45)$  \\
$F[{\Delta\phi_1]}$\tablenotemark{a}   &$4.00(3.97,4.08)$& $3.91(3.88,3.93)$& $3.94(3.91,3.96)$& $3.99(3.96,4.01)$& $3.96(3.94,3.97)$  \\
$F[{\Delta\phi_2]}$\tablenotemark{a} &$4.33(4.28,4.38)$& $4.37(4.35,4.40)$& $4.37(4.34,4.39)$& $4.45(4.42,4.47)$& $4.40(4.38,4.41)$  \\
$E_0$ (keV)                      &$0.79(0.76,0.81)$& $0.71(0.68,0.73)$& $0.72(0.69,0.74)$& $0.69(0.63,0.73)$& $0.72(0.70,0.73)$  \\
$\sigma$ (eV)                    &$\leq 62$        & $44(21,70)$      &       $68(45,95)$&     $133(78,195)$&       $69(52,89)$  \\
$EW[{\Delta\phi_1]}$ (eV)        &$40(24,51)$      & $77(59,89)$      &       $58(46,76)$&      $86(67,107)$&       $61(53,74)$  \\
$EW[{\Delta\phi_2]}$ (eV)        &$14(3,22)$       & $34(23,42)$      &       $42(32,57)$&       $39(26,52)$&       $31(26,39)$  \\
$\chi^2_{\nu}{\rm (DoF)}$        &$1.01(87)$       & $1.19(229)$      &       $1.05(269)$&       $0.99(250)$&       $1.20(212)$  \\
\enddata
\tablecomments{Results from simultaneous fits to \xmm\ spectra of \puppsr\
   extracted from two pulse-phase intervals, $\Delta\phi_1$, $\Delta\phi_2$, as defined in the Figure~\ref{lightcurve}.
   Groups numbers are defined in Table~\ref{spectable}.
   The blackbody temperatures and Gaussian line energy are linked
   between the two phases.  The column density is fixed at $N_{\rm H} =
   4.28 \times 10^{21}$~cm$^{-2}$, the phase-averaged value for the
   summed spectrum in Table~\ref{speclines}. Quoted uncertainties
   are $1\sigma$ for three interesting parameters.}
\tablenotetext{a}{Absorbed flux quoted for the $0.5-5.0$~keV band in units of $10^{-12}$~erg~s$^{-1}$~cm$^{-2}$.}
\label{phasetable}
\end{deluxetable*}

We also repeated this test for variability with the Gaussian
absorption-line model for the spectral feature (results not tabulated
here).  For either the emission-line or the absorption-line model, the
measured line centroids for all epochs but 2001 cluster well within
their 1$\sigma$ uncertainty (for two interesting parameters,
$\Delta\chi^2=2.3$).  The 2001 set deviates from the mean defined by
the other three sets by $\approx 2\sigma$.  In terms of percentage
deviation from the mean, the line centroid measured in 2001 was 14\%
and 8\% higher in the emission line and the absorption line model,
respectively.  We conclude that, although the deviation seems large
for the emission line model, it is not inconsistent with the expected
variance.  Further observations would be necessary to establish more
definite evidence of long-term variability.

\section{A Definitive Spin-Down Measurement for PSR J1210$-$5226}

Archival timing observations of \epsr\ spanning the years
2000--2008 were too sparse to securely determine its spin-down
rate, as described in \citet{hal11a}.  Searching
all possible parameter space for a phase-coherent,
quadratic ephemeris, we found two equally acceptable solutions,
with $\dot f = -1.243(22) \times 10^{-16}$ Hz~s$^{-1}$
and $\dot f = -7.084(22) \times 10^{-16}$ Hz~s$^{-1}$,
corresponding to $B_s = 9.9 \times 10^{10}$~G (solution 1) and
$B_s = 2.4 \times 10^{11}$~G (solution 2), respectively.  Since such
low $\dot E$ pulsars are generally very stable rotators with little
timing noise or glitch activity, it was deemed likely that one
of these is the true solution, and the other one is an alias with
an incorrect cycle count.  It is also important that no solutions with
smaller dipole $B_s$, and no spinning-up solutions, were found.

\clearpage

\begin{deluxetable*}{llrlcccr}
\tabletypesize{\scriptsize}
\tablewidth{0pt}
\tablecaption{Log of New X-ray Timing Observations of \epsr}
\tablehead{
\colhead{Mission} & \colhead{Instr/Mode} & \colhead{ObsID} & \colhead{Date} &
\colhead{Exposure} & \colhead{Start Epoch} & \colhead{Frequency\tablenotemark{a}} &
\colhead{$Z^2_1$} \\
\colhead{} & \colhead{} & \colhead{} & \colhead{(UT)} & \colhead{(ks)} &
\colhead{(MJD)} & \colhead{(s)} & \colhead{}
}    
\startdata
\chandra\ & ACIS-S3/CC &      14199 & 2011 Nov 25 & 31.0 & 55890.233 & 2.3577625(28) &  48.2 \\
\chandra\ & ACIS-S3/CC &      14202 & 2012 Apr 10 & 33.0 & 56027.637 & 2.3577709(43) &  22.2 \\
{\it XMM} & EPIC-pn/SW & 0679590101 & 2012 Jun 22 & 26.5 & 56100.537 & 2.3577687(25) &  68.3 \\
{\it XMM} & EPIC-pn/SW & 0679590201 & 2012 Jun 24 & 22.3 & 56102.752 & 2.3577621(34) &  69.0 \\
{\it XMM} & EPIC-pn/SW & 0679590301 & 2012 Jun 28 & 24.9 & 56106.490 & 2.3577636(23) & 109.6 \\
{\it XMM} & EPIC-pn/SW & 0679590401 & 2012 Jul 02 & 24.5 & 56110.918 & 2.3577626(28) &  61.4 \\
{\it XMM} & EPIC-pn/SW & 0679590501 & 2012 Jul 18 & 27.3 & 56126.553 & 2.3577640(27) &  51.9 \\
{\it XMM} & EPIC-pn/SW & 0679590601 & 2012 Aug 11 & 27.3 & 56150.408 & 2.3577637(23) &  81.8 \\
\chandra\ & ACIS-S3/CC &      14200 & 2012 Dec 01 & 31.1 & 56262.095 & 2.3577634(28) &  39.8 \\
\enddata
\tablenotetext{a}{Barycentric frequency derived from a $Z^2_1$ test.
The given uncertainty is for the $1\sigma$ confidence interval.}
\label{timinglog2}
\end{deluxetable*}

In 2011--2012 we started a new series of observations of \epsr\
with \chandra\ and \xmm\ that was designed to initiate and
maintain a unique, phase-connect timing solution of this pulsar
for the first time and eliminate the prior timing ambiguity.
Table~\ref{timinglog2} is a log of the new observations.
The instrumental setups and analysis
methods are the same as those described in Section 3 for \puppsr,
except that the \xmm\ source photons were extracted from a
$20^{\prime\prime}$ radius aperture instead of $30^{\prime\prime}$,
and the $0.5-2.5$~keV band was selected to optimize the pulsed signal.

\begin{deluxetable}{lc}
\tablewidth{0.9\linewidth}
\tablecaption{Ephemeris of \epsr}
\tablehead{
\colhead{Parameter} & \colhead{Value}
}
\startdata
R.A. (J2000)\tablenotemark{a}                 & $12^{\rm h}10^{\rm m}00^{\rm s}\!.91$ \\
Decl. (J2000)\tablenotemark{a}                & $-52^{\circ}26^{\prime}28^{\prime\prime}\!.4$ \\
Ephemeris Epoch (MJD TDB)\tablenotemark{b}     & 53562.0000006 \\
Ephemeris Span (MJD)                       & 51,549--56,262 \\
Frequency, $f$                                & 2.357763502865(65) Hz \\
Frequency derivative, $\dot f$                & $(-1.2363 \pm 0.0091) \times 10^{-16}$ Hz s$^{-1}$ \\
Period, $P$                                   & 0.424130748816(12) s \\
Period derivative, $\dot P$                   & $(2.224 \pm 0.016) \times 10^{-17}$ \\
Surface dipole dipole field, $B_s$          & $9.8 \times 10^{10}$ G\\
Spin-down luminosity, $\dot E$                & $1.2 \times 10^{31}$ erg s$^{-1}$ \\
Characteristic age, $\tau_c$                  & 302 Myr
\enddata
\tablenotetext{a}{Measured from \chandra\ ACIS-S3 ObsID 3913.
Typical uncertainty is $0.\!^{\prime\prime}6$.}
\tablenotetext{b}{Epoch of fitted minima of summed pulse profile; phase zero in Figure~\ref{pulses}}
\label{ephemeris2}
\end{deluxetable}

\begin{figure}
\centerline{
\hfill
\includegraphics[angle=270,width=0.9\linewidth,clip=]{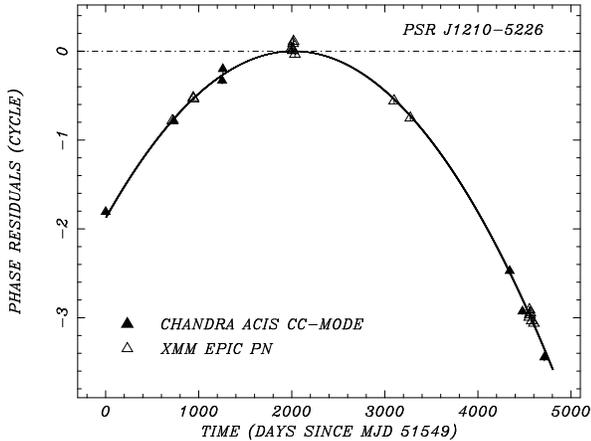}
\hfill
}
\caption{
Pulse-phase residuals from the linear term (dash-dot line) of
the phase ephemeris of \epsr\ presented in Table~\ref{ephemeris2}.
Included are the new observations listed in Table~\ref{timinglog2}
and the archival timing observations from Table~1 of \citet{hal11a}.
The quadratic term (solid line) corresponds to the uniquely determined
period derivative spanning the years 2000--2012. The error bars are
generally smaller than the symbol size.
}
\label{residuals2}
\end{figure}

\begin{figure}
\begin{center}
\includegraphics[angle=270,width=0.8\linewidth]{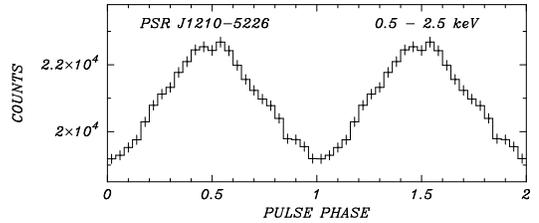}
\end{center}
\caption{Pulse profiles of \epsr\ in the $0.5-2.5$~keV band 
using data from all timing observations, folded
according to the ephemeris in Table~\ref{ephemeris2}.
Phase zero corresponds to the listed TDB epoch of the
ephemeris.  Included are the new observations in
Table~\ref{timinglog2} and the archival timing
observations from Table~1 of \citet{hal11a}.
}
\label{pulses}
\end{figure}

The coherently measured period from the new observations is precise
enough to reject previous timing solution 2, while solution 1 with the
smaller $B_s$ extrapolates precisely to the new period.  Furthermore,
the new pulse phases are aligned accurately with solution 1, while
solution 2 gives random phases.  Thus a unique quadratic ephemeris
fits all 12.9 years of \chandra\ and \xmm\ timing.
Table~\ref{ephemeris2} gives this global ephemeris, derived from a
least-square fit to the TOAs as described in Section 3, with the phase
residuals shown in Figure~\ref{residuals2}.
The previous uncertainty on the $\dot f$ of solution 1 is reduced by a
factor of 2, with the result $\dot f = -1.2363(91) \times 10^{-16}$
Hz~s$^{-1}$; the corresponding dipole magnetic field is $B_s = 9.8
\times 10^{10}$~G.  Figure~\ref{pulses} shows the pulse profile using
data from all the observations folded on the presented ephemeris.

Strictly speaking, the derived period derivative is an upper limit
to the intrinsic one, as any proper motion has not been measured
and taken into account.  \epsr\ and \puppsr\ are
at similar distances, and the location of \epsr\ with
respect to the geometric center of SNR \pks\ allows (but does not require,
since the kinematics of the SNR are unknown) a proper motion of
$\sim 70$~mas~yr$^{-1}$, or  $v_{\perp} \sim 730$~km~s$^{-1}$
\citep{del11}, similar to that of \puppsr.  If so, the Shklovskii
effect given by equation (1) would contribute
$\dot P_k \sim 1.1 \times 10^{-17}$, or half of the total $\dot P$,
and the spin-down dipole field would be reduced somewhat,
to $B_s \approx 7 \times 10^{10}$~G.  Of course, if these two CCOs
both have velocities toward the high end of the distribution of pulsars,
it would be a tantalizing physical result in itself.
However, examination of another CCO, the NS in  Cas~A, doesn't
necessarily support such high velocities for CCOs, in general;
\citet{tho01} and \citet{fes06} find $v_{\perp} \approx 350$~km~s$^{-1}$
with respect to the explosion center of Cas~A.

\section{Discussion}

\subsection{On the Age of Puppis A}

\citet{bec12} discussed the impact of the revised proper
motion of \puppsr\ on the inferred age of \psnr.  The
age of the SNR had been derived previously as $3700 \pm 300$~yr
from the proper motions of optical
filaments that point back to a common center, presumed
to be the site of the explosion, and
assuming no deceleration.  The motion of the NS also
extrapolates to the same center, but the distance traveled,
$371^{\prime\prime} \pm 31^{\prime\prime}$, corresponds to an age of
$5200 \pm 1000$~yr (or $6100 \pm 1000$~yr for our proper motion of
$61 \pm 9$  mas~yr$^{-1}$).   \citet{bec12} refer to these marginally
contradictory measurements as independent, and choose to average
them, giving the value 4.5~kyr listed in Table~\ref{ccos}.
However, they are not truly independent, because they assume
the same starting location.  Furthermore, the discrepancy
is worse if the filaments have decelerated.
Just as the new \chandra\ observations
of the NS have improved the accuracy of its proper motion,
we suggest that a contemporary observation of the optical 
filaments of \psnr\ may improve the precision of the
original \citet{win88} study on which the optical proper-motion
age is based.  Such an investigation could lead to a
more detailed understanding of dynamics of the filaments,
and a more accurate age for \psnr.

\subsection{PSR J0821--4300 and PSR J1210$-$5226 as Anti-Magnetars}

The spin-down of power \puppsr, $\dot E = 1.9 \times 10^{32}$ erg~s$^{-1}$,
is consistent with being caused by magnetic braking of an 
isolated neutron star with a weak dipole field of
$B_s = 2.9 \times 10^{10}$~G.
Its spin-down power is much smaller than its observed thermal
X-ray luminosity, $L_x \approx 5.6 \times 10^{33}\,d_{2.2}^2$~erg~s$^{-1}$,
which rules out rotation as a significant
power source.  As discussed in \citet{hal10a} for \psr, the
very small $\dot P$ disfavors propeller spin-down
and accretion as a source of the X-rays.  Thus, residual
cooling remains the most plausible source of the X-ray
luminosity of \puppsr\ and, by extension, of all CCOs.
Given the meager spin-down power of \puppsr, we can also
discount the suggestion of \citet{rgj03} and \citet{cas06}
that structure in the \psnr\ SNR and associated \ion{H}{1}
is caused by jets emitted by the pulsar, similar to
the SS433/W50 system.  Parenthetically, if the \ion{H}{1}
morphology is not caused by the pulsar, then it does not
provide supporting evidence of its distance.

An unresolved question about \puppsr\ is the origin
of its phase-dependent, possibly variable emission or
absorption feature.  In the absence of accretion, it would be
difficult to understand how an emission line is generated,
or why it would vary in a few years.  The
indication of variability is not strong, and there is
no other evidence of accretion. Therefore, we will consider here
an absorption-line interpretation, for which there is a good
precedent in \epsr, the only isolated pulsar to show a series of strong
absorption lines in its spectrum \citep{san02,mer02a,big03}.

The spectral features in \epsr\ are now widely considered to comprise
the electron cyclotron fundamental at $E_0 = 0.7$~keV and its
harmonics.  The surface magnetic field strength where the lines
are formed is inferred to be $B \approx 8 \times 10^{10}$~G
according to equation (2) and assuming $z \approx 0.3$.
The relative strength of the harmonics is explained by
treating them as resonances
in the photospheric free-free opacity in the presence
of the magnetic field \citep{sul10,sul12}.   \epsr\ is exceptional
in having the largest known spin-down dipole magnetic
field among CCOs, now confirmed as $B_s \le 9.8 \times 10^{10}$~G.
This is only slightly larger than $B \approx 8 \times 10^{10}$~G
inferred from the spectral features, and is the first case in
which such a comparison has been made between independent 
methods of measuring surface $B$-field on an isolated pulsar.
Effects that could eliminate the already minor discrepancy include
the unmeasured proper motion, which would decrease the inferred
spin-down dipole field, and the results of numerical models
of a force-free magnetosphere, which imply a different 
spin-down law from the standard vacuum dipole expression
$$B_s = \left(3 c^3 I P \dot P \over 8 \pi^2 R^6\,
{\rm sin}^2\alpha\right)^{1/2} =\ 3.2 \times 10^{19}\,
\left(P \dot P \over{\rm sin}^2\alpha\right)^{1/2}\,{\rm G}\eqno(3)$$
such that, more accurately,
$$B_s \approx \left(c^3 I P \dot P \over 4 \pi^2 R^6\,
[1+{\rm sin}^2\alpha]\right)^{1/2} =\ 2.6 \times 10^{19}\,
\left(P \dot P \over 1+{\rm sin}^2\alpha\right)^{1/2}\,{\rm G}\eqno(4)$$
\citep{spi06}. This means that a measured spin-down rate is obtained
with $B_s$ smaller by a factor $0.58-0.82$ than the conventional
approximation $B_s = 3.2 \times 10^{19} (P \dot P)^{1/2}$~G.

In the case of \puppsr, by the same logic, its weaker inferred dipole
magnetic field of $B_s \le 2.9 \times 10^{10}$~G wouldn't naturally
account for a spectral feature at $0.7-0.8$~keV, instead predicting
$E_0 \le 0.26$~keV.  The absorption model with the cyclotron
fundamental at $E_0 = 0.46$~keV comes closer to the prediction.
However, additional variables can affect the comparison of a dipole
field inferred from spin-down with that from a cyclotron line.
First is the factor of 2 variation of dipole field strength over
the surface, with $B_p = 2\,B_s$.  Second, the actual field is not
likely to be a centered dipole, and may be larger than $B_s$ or $B_p$
in places where the line is formed.  The asymmetric distribution of surface
temperature on \puppsr\ already appears to require a complex magnetic
field geometry, such as an off-center dipole or higher multipoles.
Third, the inferred spin-down field depends on the uncertain NS mass
and radius as in equations (3) and (4).
$B_s$ scales as ${I^{1/2}\,R^{-3}}
\propto M^{1/2}R^{-2}(1+z)$ \citep{rav94}, where 
$1+z = (1-2GM/Rc^2)^{-1/2} \propto B/E_0$ from equation (2).
For an astrophysically likely $M=1.4\,M_{\odot}$, theoretical
NS equations of state allow $8<R<15$~km, therefore
$1.18<1+z<1.44$.  The spectroscopic $B$ is therefore uncertain by $\pm 10\%$
when we adopt $z=0.3$, while the dipole $B_s$ is
uncertain by the much larger factor of $\sim 2$.
(The gravitational redshift cancels out in their ratio.)
Allowing for these uncertainties, it is reasonable to adopt
the hypothesis that feature(s) in the spectrum of \puppsr\ are due to
the cyclotron process.

The continuum X-ray spectrum and pulse profiles of \puppsr\
are indicative of antipodal hot spots of different areas and
temperatures \citep{got09,got10}, which are difficult
to account for if the magnetic field is weak.  A related problem
is the high pulsed fraction of 64\% from \psr\ \citep{hal10a},
a timing twin of \puppsr.
Polar cap heating by any magnetospheric accelerator must
be negligible as a source of surface heating in CCOs,
being only a small fraction of the
already insignificant spin-down power.
Thermal X-rays from residual cooling can be
nonuniform if there is anisotropic heat conduction in the star.
The effect of different magnetic field configurations
on heat transport in the crust and envelope of NSs has
been modelled, most recently by \citet{gep04,gep06},
\citet{per06a}, and \citet{pon09}.
A toroidal field is expected to be
the initial configuration generated by differential 
rotation in the proto-neutron star dynamo \citep{tho93}.
One of the effects of crustal toroidal field is to
insulate the magnetic equator from heat conduction, resulting
in warm spots at the poles.  The warm regions can even
be of different sizes due
to the antisymmetry of the poloidal component
of the field \citep{gep06}, which is evocative
of the antipodal thermal structure of \puppsr.
To have a significant effect
on the heat transport, the crustal toroidal
field strength required in these models is $\sim 10^{15}$~G,
many orders of magnitude greater than the poloidal field
if the latter is measured by the spin-down.
Purely toroidal or poloidal fields are thought
to be unstable in an initially fluid NS
\citep{tay73,flo77}, although the toroidal
field may be stabilized by a poloidal field that is several
orders of magnitude weaker \citep{bra09}.
We suggest the latter as a viable configuration for a CCO. 

\citet{sha12} tried to model the pulse profile of \psr\
with anisotropic conduction, and concluded that they
needed a toroidal crustal field of $B_{\phi} > 2 \times 10^{14}$~G
to produce its high pulsed fraction.  Even then, they
observe, the shape of the modelled light curve
doesn't match the observed one.  Since the shape of
the light curve is not reproduced, the surface
temperature distribution of \psr\ and its
physical origin are still not known.

\citet{pag96}, \citet{per06b}, \citet{zan06}, and \citet{zan07}
investigated NS surface emission patterns using
a combination of star-centered dipole and quadrupole
magnetic field components to model asymmetric pulse
profiles.  The behavior of \puppsr\ may ultimately
be explained by similar models.
It remains to be shown that if CCOs can have
field configurations that are strong enough to affect
heat transport to the extent required, while
not exceeding the spin-down limits on the
external dipole field.

\subsection{Origin and Evolution of Anti-magnetars}

\puppsr\ is nearly a twin of \psr\ in its spin properties,
and there are no other young neutron stars with measured
magnetic fields this weak.
They fall in a region of the $P-\dot P$ diagram
(Figure~\ref{ppdot}) that is devoid of ordinary (non-recycled)
radio pulsars \citep{man05}.   They overlap with the supposed
mildly recycled pulsars in this area \citep{bel10}.
The characteristic age of 254~Myr for \puppsr\
is not meaningful because the pulsar was
born spinning at its current period,
as is the case for the other CCO pulsars.
If their magnetic fields remain constant,
they will not move in $P-\dot P$ space for $>10^8$~years
(but see below).  That CCOs are found in SNRs in comparable
numbers to other classes of NSs implies that they must represent
a significant fraction of NS births. If \puppsr\ and \psr\ are
typical CCOs, the area around them in the $P-\dot P$ diagram
should be densely populated with ``orphaned CCOs'' that remain
after their SNRs dissipate in $\sim 10^5$~years.
Why there are few ordinary radio pulsars and no older X-ray
pulsars near their location is then a mystery,
as emphasized by \citet{kas10}, which may indicate
that radio luminosity is a function of spin-down power.

There are not yet enough CCOs to know whether they are
intrinsically radio-quiet relative to ordinary and
recycled pulsars, rather than unfavorably beamed.
It is also possible that some ordinary radio pulsars with $B_s < 10^{11}$~G
are actually much younger than their characteristic ages, and
may be orphaned CCOs that could be recognized in X-rays.
Whether or not they are radio pulsars, nearby orphaned CCOs
should be detectable as thermal X-ray sources for
$10^5-10^6$~years, similar to the seven \ro\ discovered
isolated NSs \citep[INSs:][]{hab07} which, however,
have strong magnetic fields \citep{kap09}. 
It is likely that the INSs are kept hot for longer
than CCOs by continuing magnetic field decay \citep{pon07}, 
which would explain their observed abundance relative
to the elusive orphaned CCOs.
It may be difficult to detect and/or
recognize orphaned CCOs if they cool faster than ordinary
neutron stars.  One effect that can accelerate cooling
is an accreted light-element envelope, which has higher
heat conduction than an iron surface \citep{kam06}.
The newly discovered 59~ms pulsar \cal\ (``Calvera''), 
originally detected in the \ro\ All-Sky Survey,
may be the first recognized example of an orphaned CCO
\citep{zan11,rut08,hal11b}, pending a measurement of its
period derivative.  The isolated NS 2XMM J104608.7$-$594306
in the Carina Nebula \citep{pir12} has been suggested as
another orphaned CCO.

It is possible that the weak magnetic field of CCOs
is causally related to slow rotation at birth through the
turbulent dynamo \citep{tho93} that generates the magnetic field.
(See \citealt{spr08} for a review of possible mechanisms for the
origin of magnetic fields in neutron stars.)  If the dipole
magnetic field $B_s$ is simply related to the initial spin period,
and assuming that the present period $P$ is in fact the birth
period because the spin-down time is so much longer than
the true age, we may expect an anti-correlation between
$B_s$ and $P$.   With only three data points to compare,
and with two of them having nearly identical
values, there is not much evidence to examine for a trend.
In fact, the CCO with the longest period,
\epsr, also has the strongest dipole field of the three,
which would not by itself support such a simple correlation.
A population analysis of radio pulsars by \cite{fau06} concludes
that there is a wide distribution of birth periods, with a
mean of $\sim 300$~ms and a dispersion of $\sigma\sim 150$~ms.
If this is true,
the birth periods of CCOs are not in fact long, and their weak
dipole fields may be the effect of some as-yet unknown parameter.

In an alternative theory for CCOs, a normal ($\sim 10^{12}$~G)
magnetic field is buried in the core or crust of a NS by prompt
fall-back of supernova debris, and takes thousands of years to diffuse
back to the surface, during which time the NS appears as an
anti-magnetar.  This is assuming that the accreted matter is itself not 
magnetized.  The timescale for diffusion is highly dependent
on the amount of matter accreted.  According to
\citet{mus95}, for accretion of $\sim 10^{-5}\,M_{\odot}$,
the regrowth of the surface field is largely complete after $\sim 10^3$~yr,
but if $> 0.01\,M_{\odot}$ is accreted, then the
diffusion time could be millions of years.
\citet{che89} calculated that the neutron star in SN~1987A could
have accreted $\sim 0.1\,M_{\odot}$ of fallback material in the hours after the
SN explosion, aided by a reverse shock from the helium layer
of the progenitor.  If so, it may never emerge as a radio pulsar.

Interesting support for the theory of field burial and regrowth is the
absence of evidence for magnetic field strengths $<10^{11}$~G in accreting
high-mass X-ray binary pulsars, as inferred from their pulse periods
and period derivatives \citep{pop13}.  If intrinsic fields in the range
$10^{10}-10^{11}$~G were common, equilibrium spin periods of $0.1-1$~s
should be frequent in HMXBs, but they are not.  If the NSs in these
systems are born in the same way as isolated pulsars, this could imply
that birth fields are never so small; instead, field regrowth has occurred
in all cases after it was initially buried.   One caveat here is that
$\sim 40\%$ of HMXBs do not have measured spin periods, although there
is no strong selection effect against detecting $P<1$~s.

\citet{ber10} has revisited the process of
hypercritical accretion onto a magnetized neutron star, while
\citet{ho11} made new calculations of the subsequent diffusion
to constrain the magnetic fields of CCOs at birth and the accreted
mass, finding $10^{-4}-10^{-5}\,M_{\odot}$ for the latter.
These models are difficult to test using CCOs, because it would involve
measuring the braking index or the change of the dipole magnetic field
directly.  \citet{ho11} implied that the isolation of CCOs
in the $P-\dot P$ diagram may be less problematic in this model
because, as their dipole magnetic fields increase, they will evolve
to join the bulk of the pulsar population.  However, for the
first $\sim 10^5$~yr, field growth can only move a CCO
vertically upward in the $P-\dot P$ diagram, as the braking index
has a large, negative value.  Orphaned CCOs should still
have periods of $\sim 0.1$~s and lie in a sparse region.
For this reason, an important test for an example of
field growth is to determine the $\dot P$ of Calvera,
which will reveal if its dipole field is
greater than that of \puppsr\ and \psr.

Previous calculations of field burial and re-emergence for CCOs
were one dimensional.
The first two-dimensional calculations for this purpose were recently
reported by \citet{vig12}.  They find that the accretion must be
essentially isotropic to bury the field sufficiently to cause
the required orders-of-magnitude reduction of the external dipole.
If the accretion were instead confined to the equator or the magnetic
poles, the reduction of the dipole component would not be significant
enough to produce a CCO.  Finally, it remains to be seen if the
resulting surface thermal distribution during the CCO phase can be
made compatible with observed spectra and pulse profiles.
\citet{vig12} briefly presented the temperature distribution from
one of their models that produces hot polar caps at a time when the
external dipole field is $10^{10}$~G.  But even in this
case, the temperature does not vary by as much as a factor of 2
over the surface, a range that would be required to match the properties
of \puppsr\ and other CCOs.

\section{Conclusions and Future Work}

Measurement of the spin-down rate of the 112~ms \puppsr\ in \psnr\
was achieved using X-ray observations coordinated between
\xmm\ and \chandra, the resulting phase-connected ephemeris
spanning 2.3 yr.  We also measured the proper motion of
the pulsar in \chandra\ HRC images over 10.6 years.
The proper motion makes a non-negligible contribution to
the period derivative via the Shklovskii effect, and the
uncertainty in proper motion and distance limits the
accuracy of the intrinsic period derivative to
$\approx 16\%$.  The straightforward interpretation
of these results is dipole spin-down due to 
a weak surface magnetic field of $B_s = 2.9 \times 10^{10}$~G,
the smallest measured for any young neutron star, and nearly
identical to that of the CCO pulsar \psr\ in \snr.

A phase-dependent spectral feature in \puppsr\ can be modelled
either as an emission line of energy $\approx 0.75$ keV, or as a 
cyclotron absorption line and its harmonic, with $E_0 \approx 0.46$~keV.
For reasons that are not clear, it is stronger during the pulse-phase
interval in which the continuum spectrum is softer.
There is only marginal evidence for long-term variability of this
feature.  The local magnetic field strength in the area where
the spectral feature is produced may have to be larger
than the dipole spin-down value.

The existing spin-down measurements for three CCOs, including a
definitive new result for \epsr, are compelling evidence that a weak
dipole field component is the physical origin of the CCO class in general.
It is reasonable to assume that the CCOs which have not yet been seen to
pulse have magnetic fields that are similar to or weaker than those of
\puppsr\ and \psr.  Otherwise, if their fields were stronger, their
spectra should show cyclotron lines similar to \epsr\ and (possibly)
\puppsr.  Deep X-ray timing searches of the remaining members of this
class, to smaller limits on pulsed fraction, could still discover
new pulsars and supply valuable data on their birth properties
and evolution.

A remaining theoretical puzzle about CCOs is the origin
of their surface temperature anisotropies, in particular, the one
or two warm/hot regions that are smaller than the full neutron star
surface.  The measured spin-down power is too small
to contribute to this emission.  Continuing
accretion is unlikely because of the small spin-down rates.
It appears that any explanation will have to involve
stronger magnetic field components in the crust,
with toroidal or quadrupolar geometries, that do
not contribute to the dipole spin-down torque.
A physical model of thermal and magnetic field
structure that self-consistently reproduces
the spin-down rates, X-ray spectra, and pulse
profiles, is still needed.  If it can be studied with better data,
the details of the phase-dependent and possibly variable
spectral feature from \puppsr\ may contribute to a more
definite model of its surface magnetic field geometry.

The CCO pulsars \psr\ and \puppsr\
fall in a region of $B-P$ space that overlaps with what are
assumed to be moderately recycled pulsars, but is otherwise empty.
An understanding of the evolutionary status of CCOs is critically
dependent upon a search for their descendants, the orphaned CCOs
without SNRs.  Some single radio pulsars with spin parameters
similar to CCOs may be orphaned CCOs rather than recycled pulsars,
and they may be much younger than their characteristic ages.
X-ray observations of pulsars in this region may
find evidence of their relative youth via surface thermal emission.
The spin-down rates of orphaned CCOs would reveal whether CCOs
have intrinsically weak magnetic fields, or if field re-emerges on a
timescale of $\sim 10^4$~yr from a normal magnetic field that was
buried by prompt fallback of supernova debris, as is invoked in some
theoretical studies.  Such rapid evolution would lead to orphaned CCOs
that lie directly above CCOs on the $P-\dot P$ diagram but are still
detectable as thermal X-ray sources.  The radio-quiet pulsar
Calvera is a possible such candidate.  Whether or not CCOs have buried
fields, the paucity of radio pulsars with similar spin parameters
is real and requires an explanation.

\acknowledgments
We thank Dany Page, and an anonymous referee, for helpful comments.
This investigation is based on observations obtained with \xmm,
an ESA science mission with instruments and contributions directly
funded by ESA Member States and NASA, and \chandra.
The opportunity to propose joint, coordinated observations
between \xmm\ and \chandra\ was crucial for the success of
this effort.  Financial support was provided by NASA grants
NNX11AD19G and NNX12AD41G for the \xmm\ observations,
and by {\it Chandra} awards GO1-12001X and
SAO GO1-12071X, issued by the \chandra\ X-ray Observatory Center,
which is operated by the Smithsonian Astrophysical Observatory
for and on behalf of NASA under contract NAS8-03060.

\end{document}